\documentclass[11pt,onesidepages]{article}

\usepackage{enumerate}
\usepackage{amsmath}
\usepackage{multicol}
\usepackage{tikz}
\usepackage{subfigure}
\usepackage{cases}
\usepackage{cite}
\usepackage[square, numbers, comma, sort&compress]{natbib}
\usepackage{balance}

\newcommand{\virgolette}[1]{``#1''}

\begin{document}

\title{Design of Graded Photonic Crystals Antennas via Inverse Scattering based techniques}

\author{Roberta~Palmeri,
        Martina~T.~Bevacqua, \\Andrea~F.~Morabito,
        and~Tommaso~Isernia
\thanks{All the authors are with Dipartimento di Ingegneria dell'Informazione, delle Infrastrutture e dell'Energia Sostenibile, Universit\`a \virgolette{Mediterranea} di Reggio Calabria, via Graziella, Loc. Feo di Vito, 89124, Reggio Calabria (Italy). }
}

\maketitle

\begin{abstract}
A new approach to the design of graded Photonic Crystals (GPCs) devices is proposed by exploiting the inverse scattering framework as a synthesis tool. The introduced general methodology can be applied to arbitrary far-field specifications, thus allowing to design non-canonical devices. In particular, two different strategies are developed which allow to deal with GPCs with both graded refractive index ($GPC_R$) and graded filling factor ($GPC_F$). In both strategies, the inverse
scattering problem is solved by a proper reformulation of the Contrast Source Inversion method wherein a proper rescaling of the amplitude of the primary sources is pursued. In particular, in the first one, the GPCs are obtained by exploiting homogenization theories. In the second one, the $GPC_R$ profile is synthesized by exploiting a suitable representation basis for the unknown contrast function and, then, simple analytical formulas are used to determine a $GPC_F$.
The proposed approach is assessed through the synthesis of an antenna generating $\Sigma/\Delta$ reconfigurable patterns.
\end{abstract}

\section{Introduction}

Graded Refractive INdex (GRIN) media  allow to control the electromagnetic field paths. Classical GRIN devices are the Luneburg lens \cite{morgan1958general,luneburg1964mathematical} and the Maxwell fish-eye lens \cite{fisheye} (with their variants \cite{aghanejad2012design,duan2008enhancement,jiang2008layered,tichit2009ultradirective,loo2012broadband}). These elegant solutions of the Maxwell equations allow to design and build useful antennas and devices. On the other side, they do not allow to enforce a different given behavior of the electromagnetic field.
To this end, more flexible techniques have been introduced in literature. 

As a first possibility, the Transformation Optics (TO) theory \cite{pendry2006controlling} gives a general methodology for controlling the electromagnetic waves propagation by tailoring the spatial constitutive profile of a material \cite{aghanejad2012design,duan2008enhancement,jiang2008layered,tichit2009ultradirective,gallina2008transformation,moccia2016dispersion}. 
Unfortunately, the bi-anisotropic materials generally obtained by TO are so complicated that cannot be easily realized. So, in order to have more easily manufacturable optical devices, one can relax the exact required parameters of the material, by paying the price of an unavoidable deterioration of the performances.

A suitable and more flexible alternative is represented by the inverse scattering theory, which provides an interesting framework for the synthesis of dielectric profile antennas \cite{bucci2005synthesis,isernia2017aps}, as well as of other devices \cite{didonato2014cama,didonato2017cloaking}. In fact, it represents a general methodology which allows to control the electromagnetic waves path in order to obtain arbitrary far field specifications.

Inverse scattering techniques \cite{pastorino,colton2012} are nowadays widely exploited for microwave imaging in biomedical diagnosis, remote sensing, non-destructive testing, and so on. \\
In an inverse scattering problem (ISP) the aim is to retrieve location, shape and electromagnetic properties of an unknown object starting from the knowledge of the incident field and the measurements of the arising scattered or total field. 
Such diagnosis problems can be turned into synthesis problems by considering a specific behavior of the total field as available data of the problem (rather than the measured total field). In this case, the new aim of the problem is to determine the dielectric profile of an object such that the interaction with an impinging wave will give rise to the specified total field. 

Notably, the solution of an ISP represents a very difficult task as it is non linear and one can run into the so-called \virgolette{false solutions} \cite{isernia2001local}, which could be completely different from the actual ground truth. However, in the adoption of the ISP as a design tool,  the non linearity of the inverse scattering problem is not equally problematic, since whatever dielectric profile is admissible as long as a good matching with the expected field characteristics is achieved. 

From a practical point of view, the realization of a GRIN lens with a generic gradient index profile poses difficult fabrication challenges. Hence, investigations have been performed on stepped-index lenses \cite{mosallaei2001nonuniform,fuchs2007design,fuchs2006design}, in which the desired continuous variation of index with radius is approximated by a number of constant-index spherical shells. As it can be easily guessed, such a strategy results in a tradeoff between the number of shells and the achieved performances.

Recently, more effective fabrication techniques have been developed based on the use of graded Photonic Crystals (GPCs)\cite{joannopoulos2011photonic,centeno2005graded,gaufillet2012graded,vasic2010controlling}, thanks to which the guiding of the electromagnetic waves is performed using a well-designed spatially dependent dispersion, i.e., by engineering the filling factor, the lattice period, and/or material index. Such structures are widely used in literature and good results have been obtained for the realization of canonical lenses \cite{loo2012broadband,sun2014luneburg,zhao2016anisotropic,gaufillet2016graded,liu2016luneburg,falek2015parametric}. Note that it is possible also to tune these kinds of devices by exploiting, for instance, electro-optic or thermo-optic effects \cite{levy2008tunable,markos2010bending}. \\
The design of the above explained GPCs is essentially based on the homogenization theory, among which the Maxwell-Garnett effective medium \cite{datta1993effective} is the most frequently used. 

In this paper, we introduce a general methodology which can be applied to arbitrary far field specifications and allows to design two-dimensional GPCs by exploiting the inverse scattering framework. 
Beside being a step towards the realization of 3D PC-based antennas, consideration of 2D devices has an interest \textit{by per se}. In fact, a practical implementation can be faced by using parallel plate waveguides \cite{wiesbeck2002,sato2002, xue2008}, so that a proper `arraying' along the residual direction can allow actual 3D design solutions. Moreover, the 2D problem herein considered shows interesting relationship with the synthesis of flat antennas, whose interest is recently growing \cite{pfeiffer2010printed,bosiljevac2012luneburg,gonzalez2017,pavone2017liquid}. In fact, in both cases one has to realize a proper arrangement of the field paths along a plane. 

The overall procedure consists in two steps. In the first one, a far field pattern obeying to some mask constraints is synthesized. Then, in the second step, the ISP is solved in order to synthesize a device able to radiate the far field pattern resulting from the previous step. Since the final goal of the approach is to design a GPCs device with a gradient of the filling factor, we propose and compare two different strategies to accomplish such a goal. In particular, a first and more intuitive possibility concerns the use of the effective medium theory on the obtained continuous GRIN profile.
As a second and more original possibility, we propose an approach based on the use of a convenient basis expansion of the contrast function, which encodes the e.m. properties of the unknown target, and simple analytical arguments. 

The paper is structured as follow. In Section \ref{sec:inv_prob} the basics of the ISP and the Contrast Source Inversion (CSI) \cite{vandenberg1997,vandenberg1999extended,isernia1997nonlinear} method are recalled. In Section \ref{sec:inv_prob_sint} a reformulation of the CSI is given allowing to tune the amplitude of the primary sources, while the GPCs synthesis procedures are explained in Section \ref{sec:inv_proc}. Finally, in Section \ref{sec:example} the proposed approach is assessed through the design of an antenna for radar applications, while details about the synthesis of the far field patterns of the specific example at hand are given in the Appendix. Conclusions follow.

\section{The inverse scattering problem and the Contrast Source Inversion method} 
\label{sec:inv_prob}

Let us consider a region of interest $\Omega$ in which the background medium is air ($\varepsilon_b=1$, $\sigma_b=0$ being the relative permittivity and conductivity, respectively), and let $\Gamma_{t}$  and $\Gamma_{o}$ be the illumination and the observation surface, respectively.

For a 2D TM scalar problem (and omitting the implicit time harmonic factor $\exp(j\omega t )$), a rather usual formulation of the ISP \cite{pastorino,colton2012} reads:

\begin{equation}\label{eq_data}
\begin{split}
E_s(\textbf{r}_o,\textbf{r}_t)= \mathcal{A}_e[W(\textbf{r},\textbf{r}_t)] \\
= -\frac{j\beta_0^2}{4} \int_\Omega H_0^{(2)}(\beta_0 |\textbf{r}_o-\textbf{r}'|) W(\textbf{r}',\textbf{r}_t) d\textbf{r}'  , \ \ \ \textbf{r}\in\Omega
\end{split}
\end{equation}

\begin{equation}\label{eq_state}
\begin{split}
W(\textbf{r},\textbf{r}_t)-\chi(\textbf{r})E_i(\textbf{r},\textbf{r}_t) = \chi(\textbf{r})\mathcal{A}_i[W(\textbf{r},\textbf{r}_t)] \\
=-\frac{j\beta_0^2}{4}\chi(\textbf{r}) \int_\Omega H_0^{(2)}(\beta_0 |\textbf{r}-\textbf{r}'|) W(\textbf{r}',\textbf{r}_t) d\textbf{r}' , \ \ \ \textbf{r}\in\Omega
\end{split}
\end{equation}
where $E_i$ and $E_s$ are the incident and scattered field, respectively, $W=E_t\chi$ is the so called contrast source, and $E_t=E_i+E_s$ is the total field. The kernel $H_0^{(2)}(\beta_0\textbf{r})$ of the integral operator is the Hankel function of zero order and second kind, $\beta_0$ is the wavenumber in the background medium, $\textbf{r}_t=(r_t,\theta_t)\in\Gamma_t$  and $\textbf{r}_o=(r_o,\theta_o)\in\Gamma_o$ are the illumination and the observation position, respectively, while $\mathcal{A}_e$ and $\mathcal{A}_i$ are a short notation for the integral radiation operators. 

The aim of the ISP is to retrieve the contrast function $\chi$ starting from the knowledge of the scattered field $E_s$ on $\Gamma_{o}$. The contrast function \begin{math} \chi(\textbf{r})=(\tilde{\varepsilon}_r(\textbf{r})-\varepsilon_b)/\varepsilon_b\end{math} encodes the electromagnetic properties of the region $\Omega$; in particular, $\tilde{\varepsilon}_r(\textbf{r})=\varepsilon_r(\textbf{r})-j\sigma_r(\textbf{r})/\omega\varepsilon_0$ is the complex relative permittivity, $\omega$ being the angular frequency and $\varepsilon_0$ the permittivity of the vacuum.

As is well known, the ISP is non linear since both the contrast function $\chi$ and the contrast source $W$ are unknowns \cite{pastorino,colton2012}. In order to face such a difficulty, several efforts have been carried out in the literature to develop effective solution methods \cite{vandenberg1997,vandenberg1999extended,isernia1997nonlinear,isernia2001local,abubakar2001total}. The contrast source inversion (CSI) method \cite{vandenberg1997,vandenberg1999extended} is one of the most popular and effective inversion schemes. As a matter of fact, it allows to face the ISP in its full non-linearity, while dealing with a mathematical problem involving just linear and quadratic equations. In particular, it simultaneously looks for both the contrast $\chi$ and the contrast source $W$, and the solution is iteratively built by minimizing the cost functional (\ref{func1}), which takes into account the data-to-unknown relationship and the physical model \cite{vandenberg1997,vandenberg1999extended,isernia1997nonlinear,isernia2001local}: 
\begin{equation}\label{func1}
\begin{split}
\Phi(W,\chi)=\sum_{v=1}^T \frac{  \left \| \mathcal{A}_e \left [W^{(v)} \right ] - E_s^{(v)} \right \|_2 ^2 }{ \left \| E_s^{(v)} \right \|_2 ^2 } \\
 + \sum_{v=1}^T \frac{\left \| \chi\mathcal{A}_i \left [W^{(v)} \right ] +\chi E_i^{(v)} -W^{(v)} \right \|_2 ^2 }{\left \| E_i^{(v)} \right \|_2 ^2 },
\end{split}
\end{equation}
where $\left \| \cdot \right \|_2$ is the $\ell_2$-norm and $T$ is the number of different incident fields corresponding to different position $\textbf{r}_t$. 

As proposed in \cite{isernia1997nonlinear}, the minimization of (\ref{func1}) can be pursued by means of a conjugate gradient algorithm in which, at each step, the values of $\chi$ and $W^{(v)}$ are updated by a line minimization procedure. 
Obviously, other minimization schemes (including off the shelf numerical routines) can  be used. In fact, this will not affect the final results, the only resulting difference being the computational time.

\section{A modified CSI as a design tool} 
\label{sec:inv_prob_sint}

When a synthesis problem is considered instead of a diagnostics one, the aim becomes determining $\chi(\textbf{r})$ (i.e., the electromagnetic properties of the $\Omega$ region)  starting from the knowledge of the incident field and obeying given specifications of the total field. Let us note explicitly that design constraints are in terms of the total fields, which is slightly different from more usual ISP, where equations are usually written in terms of the scattered field. Obviously, one can easily go from one formulation to another by simply subtracting or adding the incident fields.

Then, it is also worth to note that for a given total field on the observation domain $\Gamma_o$, different amplitudes of the incident fields give rise to different requirements on the scattered fields (and hence to different profiles). For this reason, it proves convenient to modify the standard CSI algorithm by considering one more set of complex unknowns $\tau^{(v)}$ modulating the amplitudes of the primary sources. In fact, exploitation of these additional degrees of freedom will allow for a better matching of the desired fields, and/or to simpler contrast profiles. Then, it is convenient to distinguish among a `basic' incident field $\widehat{E}_i^{(v)}$ (corresponding to unitary excitation) and on `actual' incident field $E_i^{(v)}$ (corresponding to the synthesized excitations of the primary sources times the corresponding $\widehat{E}_i^{(v)}$).\\
From the above, we recast the CSI functional as follow:
\begin{equation}\label{func}
\begin{split}
\Phi(W,\chi,\tau)=\sum_{v=1}^T \frac{  \left \| \mathcal{A}_e \left [W^{(v)} \right ] - E_t^{(v)} +\tau^{(v)} \widehat{E}_i^{(v)} \right \|_2 ^2 }{ \left \| E_t^{(v)} \right \|_2 ^2 } \\ + \sum_{v=1}^T \frac{\left \| \chi\mathcal{A}_i \left [W^{(v)} \right ] +\chi \tau^{(v)} \widehat{E}_i^{(v)} -W^{(v)} \right \|_2 ^2 }{\left \| \widehat{E}_i^{(v)} \right \|_2 ^2 } ,
\end{split}
\end{equation}
where we split the scattered field as \begin{math} E_s=E_t-\tau \widehat{E}_i\end{math}, $E_t$ being in this case the given total field on $\Gamma_o$, i.e., our design constraint.

As it can be seen, the normalization herein considered for the first addendum is not the same of functional \eqref{func1}; in fact, the total field is considered instead of the scattered field since this latter changes its value at each iteration due to the rescaling of the incident field by $\tau^{(v)}$.

The solution of the inverse scattering design problem can still be solved by minimizing the cost functional \eqref{func} and by adopting the procedure developed in \cite{isernia1997nonlinear}.

A simple modification of the proposed synthesis tool (see below) concerns the possible addition of penalty terms to the cost functional in order to enforce some desired behavior on the profile. 
In fact, besides the total field specifications, one could require some desired properties also on the contrast function. Notably, the additive penalty terms enforce the desired behavior on $\chi$, while the original term $\Phi(W,\chi,\tau)$ penalizes the violation of the data and physical model mismatch \cite{isernia1997nonlinear,didonato2015csive,bevacqua2017csquadro}.
In summary, the final optimization problem reads: 
\begin{equation}\label{func_pen}
\min_{W,\chi,\tau} \Phi' = \min_{W,\chi,\tau} \left( \Phi + k_p \Phi_p \right) ,
\end{equation}
where $k_p$ is a positive weighting coefficient of the occurring penalty term at hand. If it is sufficiently large, the minimization is enforced to evolve inside or close to the set implicitly defined from the meant constraints. More details on the choice of $k_p$ are given in the numerical section.

A first possible requirement on $\chi$ (which is of interest in the following) could be enforcing a circular symmetry. In this case, the additive penalty term can be expressed as:
\begin{equation}\label{pen_circ}
 \Phi_p=\Phi_s=\left \| \frac{\partial \chi}{\partial \theta}  \right \|_2 ^2 ,
\end{equation}
since the minimization of $\Phi_s$ allows to minimize the angular variation of the contrast function around the center of the coordinate system.

A second useful requirement could be enforcing lossless and physical feasibility properties of the contrast function in order to possibly avoid metamaterials. Obviously, this is not strictly needed, but such a property can allow an easier manufacturing of the device. 
To this aim, the pertaining additional term can be formulated as follow: 
\begin{equation}\label{pen_fis}
 \Phi_p=\Phi_f=\left \| \chi - f( \chi )  \right \|_2 ^2 ,
\end{equation}
where $f( \chi )$ is the projection of $\chi$ into the set of admissible functions (e.g., the set of real and positive functions) \footnote{It is worth to note that by \virgolette{physically feasible} we simply mean herein that the permittivity values are larger than 1, so that `natural' materials could be eventually used.} .

Finally, in case of circularly symmetric profiles, an interesting change for simplified manufacturing occurs in presence of a reduced number of materials when moving along the radial coordinate. Such a property is strictly related to the basic `sparsity' concept of the Compressive Sensing (CS) framework \cite{donoho2006compressed}. By deferring to \cite{donoho2006compressed} for more details, CS theory and sparsity promoting techniques \cite{winters2010sparsity,ambrosanio2015compressive,shah2016inverse,palmeri2017microwave,bevacqua2017csquadro,isernia2017aps} are of interest herein, as one proves that the minimization of the $\ell_1$-norm of the radial derivative of $\chi$ both enforces a piecewise constant behavior on the contrast profile and promotes the minimal number of hops. In such a case, the arising penalty term reads \cite{bevacqua2017csquadro,isernia2017aps}:
\begin{equation}\label{pen_spars}
 \Phi_p=\Phi_r=\left \| \frac{\partial \chi}{\partial r}  \right \|_1 ,
\end{equation}
denoting by $\left \| \cdot \right \|_1$ the $\ell_1$-norm. 

As a final comment, note that the reformulation of the cost functional \eqref{func} (as well as the addition of penalty terms) leads to a modification of the gradient of the functional and of the coefficients in the line minimization step as well. For brevity, we do not provide the new expressions since they can be easily calculated following the procedure in \cite{isernia1997nonlinear} for the functional  and \cite{didonato2015csive,bevacqua2017csquadro} for the penalty terms. 

\section{Graded Photonic Crystals (GPCs) design tools}\label{sec:inv_proc}

The practical realization of GRIN structures is not a trivial task due to the need of realizing arbitrary index gradients in a controlled manner. To overcome this drawback, a first approach looking towards a manufacture simplification could be an a-posteriori discretization of the GRIN profile by means of a number of shells \cite{isernia2017aps}. In this case, some optimization procedures can be developed to reduce the foreseen degradation of the performances \cite{mosallaei2001nonuniform,fuchs2007design,fuchs2006design}.

In the last decades, a great interest has been devoted to realize GRIN structures by means of GPCs \cite{centeno2005graded,gaufillet2012graded,vasic2010controlling}, since a suitable engineering of the basic structure allows to control the electromagnetic waves propagation. In particular, such a property can be achieved thanks to a gradient of the refractive index or to a gradient of the filling factor; hereinafter, we will refer to $GPC_{R}$ for the former structure and to $GPC_{F}$ for the latter.

In the following, two different approaches are proposed to obtain a GPCs device by taking advantage from the introduced inverse scattering methodologies. Note in both cases we pursue the realization of a $GPC_{F}$. 

A first, quite straightforward, strategy amounts to exploit homogenization procedures on the solution of the inverse scattering problem (Section \ref{sec:MG}).
The second, and more sophisticated, strategy is based instead on an original representation of the contrast function followed by analytically based equivalences among different small scatterers (see Section \ref{sec:rods}). \\
In order to introduce and validate the proposed tools in a simple yet significant case, we deal in the following with a circularly symmetric profile. Notably, such a choice allows steering the beam (without incurring into any performance deterioration) by accordingly moving the primary sources. Moreover, in both cases, we exploit \cite{sun2014luneburg} in order to define the geometrical structure of the GPCs. In particular, the adopted unit cell is a triangle, so that the overall GPCs structure exhibits six-fold rotational symmetry.

\subsection{Case 1: a Maxwell-Garnett driven procedure}\label{sec:MG}
The Maxwell-Garnett (MG) effective medium theory \cite{datta1993effective} is widely used to define the effective permittivity of the GPCs structures with rods embedded in a host medium. Therefore, if the arrangement of the rods is known and a polarization for the field is chosen, by applying the MG mixing formulas \cite{sihvola1999} to a GRIN profile, the local permittivity value is realized by means of a proper choice of the radii of the different rods (which are made by the same material). This lead to a $GPC_{F}$ structure and such an approach has been successfully applied in literature in case of canonical GRIN devices \cite{loo2012broadband,sun2014luneburg,zhao2016anisotropic,gaufillet2016graded,liu2016luneburg,falek2015parametric}. 

The strategy proposed in this subsection applies the above described homogenization theory to the continuous GRIN profile obtained by solving the ISP (wherein a symmetric behavior has been enforced by means of penalty term $\Phi_s$ in eq. \eqref{pen_circ}). To this end, the GRIN profile is sampled in a sufficiently dense number of points corresponding to the location of the different rods. Finally, once the collocation of the rods is determined and a convenient material is chosen, the mixing formulas are applied to obtain the equivalent $GPC_{F}$.

\subsection{Case 2: synthesis of GPCs}\label{sec:rods}
As an alternative to the GRIN-to-GPCs transition by means of homogenization procedures, is this subsection an approach that allows to directly look for a GPCs structure is introduced and discussed. In order to accomplish such a goal, the unknown contrast function is expanded by means of a proper basis function which projects it into the `space of rods', i.e.:
\begin{equation}\label{chi_basis}
\chi (\textbf{r}) = \sum_{k=1}^K \chi_k \sum_{h=1}^{H_k}  \Pi \left ( \frac{\textbf{r}-\textbf{r}_{kh}}{a_k} \right ) ,
\end{equation}
where $\chi_k$ is the contrast value associated to the $k$-th ring of rods, $K$ is the total number of the rings for the arrangement of the rods, $H_k$ is the number of inclusions along the $k$-th ring, $\textbf{r}_{kh}$ is the position of the center of the $h$-th rod belonging to the $k$-th ring, and $\left | \textbf{r}_{kh} \right | = r_k$. 
Finally, each $\Pi \left ( \frac{\textbf{r}-\textbf{r}_{kh}}{a_k} \right )$ function (which is associated to a single rod) is a circular window of radius $a_k$ centered in $\textbf{r}_{kh}$. As a consequence, the internal summation defines a composite window which is different from zero in each rod belonging to the $k$-th ring, and zero elsewhere.

By using representation \eqref{chi_basis} into minimization of functional \eqref{func}, the ultimate unknowns of the problem become the coefficients $\chi_k$, which represent the contrast values of the rods belonging to the $k$-th ring.\\
The outcome of such an optimization is a $GPC_{R}$ device. Although this result is of interest \textit{by per se}, a more interesting design solution is to determine $GPC_{F}$ devices, in which the radii (in each ring) are instead the actual degrees of freedom of the problem. However, the direct search for the radii is a very difficult task because of the indirect way the unknowns $a_k$ enter into the inverse scattering problem, thus increasing its non linearity. 

Very interestingly, by using classical analytical tools one can exploit the (partial) above result, i.e., the $GPC_{R}$ structure, to synthesize the $GPC_{F}$ device in a simple fashion. 
In particular, some smart determination of the radii can be pursued by taking advantage from
the fact that the scattering behavior of each inclusion can be conveniently analyzed in terms of a so-called scattering matrix \cite{jones1986acoustic}. More in detail, by adopting a cylindrical coordinate system centered on the axis of a circular cylinder of radius $a$, we can write an expansion in cylindrical harmonics for the incident ($\widetilde{E}_i$), total ($\widetilde{E}_t$) and scattered ($\widetilde{E}_s$) field pertaining to a single inclusion \cite{jones1986acoustic}:
\begin{equation}\label{eq:exp_ei}
\widetilde{E}_i(r,\theta)=\sum_{n=-\infty}^{+\infty} a_n J_n \left( \beta_0 r \right) e^{jn\theta} ,
\end{equation}
\begin{equation}
\widetilde{E}_t(r,\theta)=\sum_{n=-\infty}^{+\infty} b_n J_n \left( \beta r \right) e^{jn\theta} ,
\end{equation}
\begin{equation}\label{eq:exp_es}
\widetilde{E}_s(r,\theta)=\sum_{n=-\infty}^{+\infty} c_n H_n^{(2)} \left( \beta_0 r \right) e^{jn\theta} ,
\end{equation}
where $a_n$, $b_n$ and $c_n$ are the expansion's coefficients, $J_n$ and $H_n^{(2)}$ being the $n$-th order Bessel function and Hankel function of second kind, respectively, while $\beta$ is the wave number of the dielectric medium filling the cylinder. \\
Hence, the \virgolette{response} of a homogeneous cylindrical scatterer can be conveniently analyzed in terms of the scattering coefficients \begin{math}s_n=c_n/a_n \end{math}. 
By applying the recurrence formulas for the derivatives of the Bessel functions, one finds \cite{abramowitz}:
\begin{equation}\label{eq:exp_coeff}
s_n=\frac{\left( \beta a \right) J_{n-1}\left( \beta a \right) J_n\left( \beta_0 a \right) - \left( \beta_0 a \right)J_n\left( \beta a \right) J_{n-1}\left( \beta_0 a \right)   }{\left( \beta_0 a \right) J_n\left( \beta a \right) H_{n-1}^{(2)}\left( \beta_0 a \right) - \left( \beta a \right) J_{n-1}\left( \beta a \right) H_n^{(2)}\left( \beta_0 a \right)   } .
\end{equation}
Note that besides the dependence on the radius $a$, $s_n$ intrinsically depends on the contrast function $\chi$ through $\beta_0$ and $\beta$. 

\begin{figure}[htbp]
\centering
\includegraphics[scale=0.36]{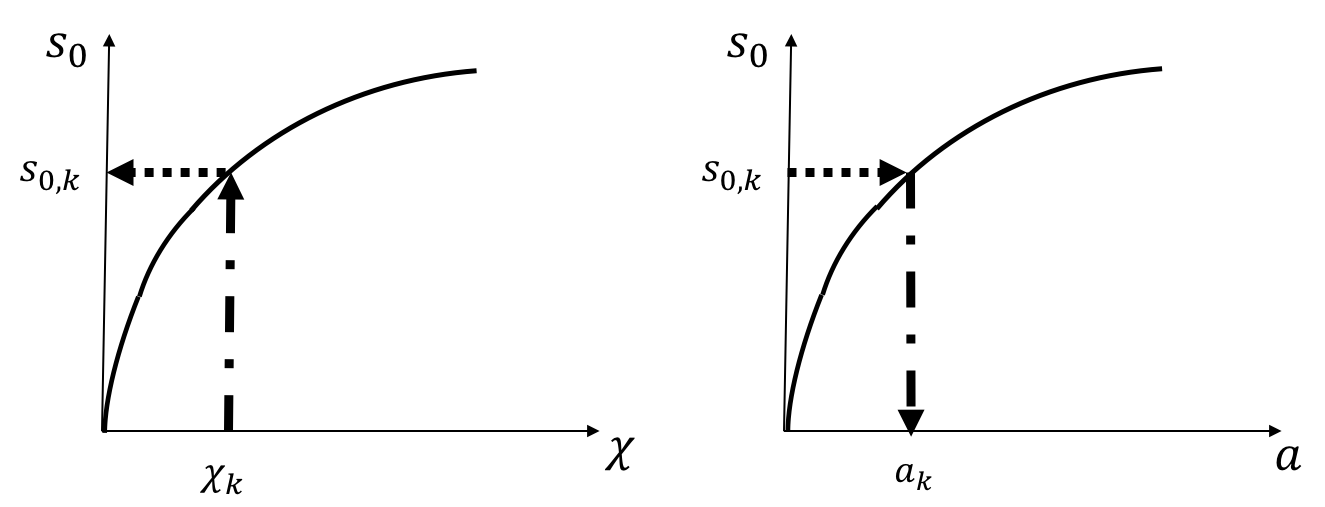}
\caption{Qualitative representation of the analytic-based tool to design the $GPC_{F}$ starting from the $GPC_{R}$ structure. Left: $s_0$ scattering coefficient for varying $\chi$ and assigned the radius of the $k$-th rod. Right: $s_0$ scattering coefficient for varying $a$ and assigned a contrast value for the rods.}
\label{fig:interchange}
\end{figure}

As long as the dielectric cylinder is sufficiently small with respect to the wavelength, one can easily verify by numerical or analytical tools that the term $s_0$ is much larger than all the others, and the scattering phenomenon is essentially determined from the $n=0$ term. Then, one can keep unaltered the behavior of $s_0$ (and in a first instance of the overall scattering phenomena) by performing an interchange between the local contrast value $\chi_k$ of the $k$-th rod and the radius $a_k$. \\
The guidelines of such an interchange are qualitatively shown in Fig. \ref{fig:interchange} and summarized in the following:
\begin{enumerate}[(i)]
\item evaluate by means of equation (\ref{eq:exp_coeff}) the value $s_{0,k}$ corresponding  to each $k$-th ring of the above obtained $GPC_{R}$;
\item evaluate $s_0$ as a function of the radius $a$ (by using the value of $\chi$ meant to realize the $GPC_{F}$);
\item determine the radius of the inclusions on the $k$-th ring (i.e., $a_k$) such to realize the value of $s_{0,k}$ as determined in (i).
\end{enumerate}

\section{Synthesis of a lens antenna generating $\Sigma/\Delta$ reconfigurable patterns}\label{sec:example}

The proposed inverse scattering based design approach is quite general and hence viable for different kinds of devices. In this paper, we assess the proposed technique with respect to the design of a lens antenna generating a $\Sigma/\Delta$ reconfigurable pattern \cite{elliot_antenna}, which is useful for monopulse radar applications. 
In the literature, a lot of methods are available for the synthesis of monopulse antennas, usually relying on either arrays or reflectors \cite{ares1996optimal,lee1993subarray,morabito2010optimal,rocca2015optimal}. With respect to common architectures, an interesting alternative could be represented by properly designed circularly symmetric dielectric lenses, since they allow to overcome beam degradation or mechanical scanning problems.

The synthesis of the lens antenna is pursued by adopting the proposed design method, which can be summarized as follows:
\begin{enumerate}[i.]
\item \textit{Definition of the \textit{far field} design constraint}: by exploiting the approach in \cite{morabito2010optimal}, and taking into account the available degree of freedom \cite{bucci1997dof}, two convenient $\Sigma/\Delta$ far field patterns are synthesized which obey to given mask constraints and goals: 
\item \textit{Determination of the equivalent \textit{near field} target fields}: in order to avoid possible numerical drawbacks which could arise when reasoning in terms of far fields, the observation domain $\Gamma_o$ is positioned in the near field region and a backpropagation (from the synthesized far fields) is used in order to evaluate the target fields $E_t$ on $\Gamma_o$;
\item \textit{Solution of the inverse scattering problem and Design of the GPCs}: once the total field $E_t$ on $\Gamma_o$ has been defined, the optimization problem involved in the modified CSI method is solved. Then, since the final goal of the approach is to design a $GPC_{F}$ device, the two alternative strategies described in Section \ref{sec:inv_proc} are applied. 
\end{enumerate}
More details about points i. and ii. are given in the Appendix.

\begin{figure*}[htbp]
\centering
\includegraphics[scale=0.35]{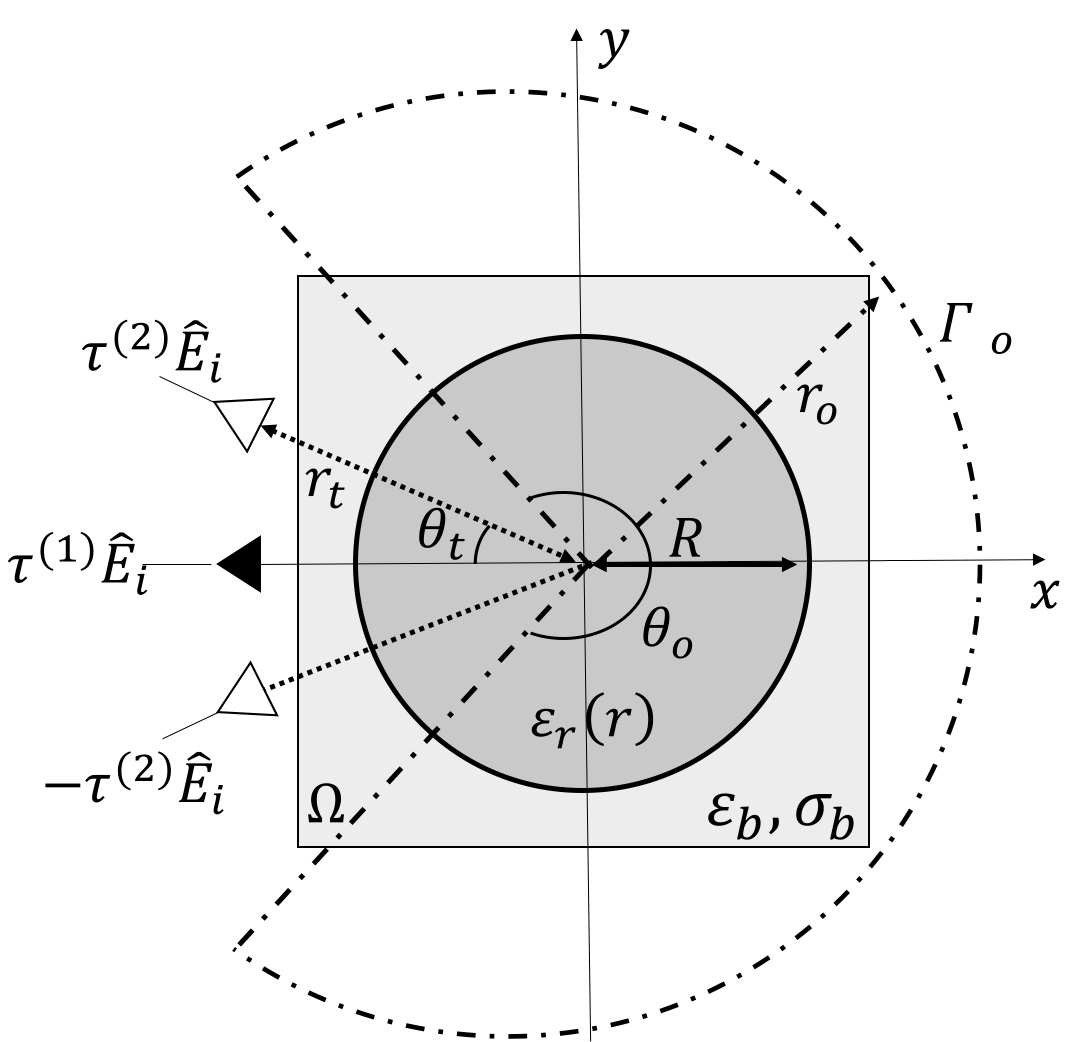}
\caption{The reference scenario for the numerical assessment. Two sets of excitations are depicted: if the \virgolette{black triangle} antenna is active, then a sum ($\Sigma$) pattern is radiated; if the \virgolette{white triangle} antennas are active and excited with an opposite phase, then a difference ($\Delta$) pattern is radiated.}
\label{fig:scenario_esempio}
\end{figure*}

In order to solve the ISP, we consider \begin{math}\widehat{E}_i^{(v)} \left( \textbf{r} \right)=H_0^{(2)}\left( \beta_0 r \right)cos^4\vartheta \end{math} as primary source, $\textbf{r}=(r,\vartheta)$ being the coordinate of the generic point belonging to a reference system centered on the phase center of the feed. In particular, the design constraints imply that $T=2$ primary incident fields have to be used. By referring to Fig. \ref{fig:scenario_esempio}, if the one placed at $\theta_t=0$ is active, the $\Sigma$-pattern is provided, while, when the two feeds located at $\theta_t>0$ are simultaneously active and excited with an opposite phase, the corresponding total field will provide the $\Delta$-pattern. In all cases, numerical codes based on the method of the moments have been exploited \cite{richmond} so that the region of interest $\Omega$ has been discretized into $N \times N$ square cells. Note that, in the procedures described in Section \ref{sec:inv_proc}, a very dense discretization has been considered in order to correctly model the small circular windows involved in representation \eqref{chi_basis} as well as in the mixing formulas and the interchanging tool. 
In order to represent the reference field in a non redundant fashion and to enforce an accurate fitting of the \virgolette{measurements} data in equation \eqref{eq_data}, a number of sampling points has been adopted as in \cite{bucci1997dof}.

In Table \ref{tab:parameters} some other useful parameters involved in the numerical simulations are reported (see also Fig. \ref{fig:scenario_esempio}).

\begin{table*}[h!]\footnotesize
\centering
\caption{\bf Key parameters of the numerical experiment.}
\renewcommand\arraystretch{1.3} 
\begin{tabular}{c|c|c}
\textbf{Parameter} & \textbf{Description} & \textbf{Value} \\
\hline \hline
$R$ & Radius of the antenna & $2\lambda$ \\
\hline
$D$ & Side of the region $\Omega$ & $4.04\lambda$ \\
\hline
$r_t$ & Distance of the primary sources from the origin of the reference system & $3.1\lambda$ \\
\hline
$\theta_t$ & Elevation angle of the primary sources  & $20^\circ$ \\
\hline
$r_o$ & Radius of the arc $\Gamma_o$ on which the design constraints are imposed & $4.5\lambda$ \\
\hline
$\theta_o$ & Central angle of the circular sector defined by the arc $\Gamma_o$  & $240^\circ$ \\
\hline
\end{tabular}\label{tab:parameters}
\end{table*}

By following the strategy of Section \ref{sec:MG}, we first achieve the continuous GRIN lens shown in Fig. 3(a). In using the modified CSI method, we added the penalty terms $\Phi_s$ and $\Phi_f$ to the cost functional \eqref{func}, in order to enforce a circularly symmetric permittivity profiles, and permittivity values real and larger than 1.
In fact, we enforced conductivity values $\sigma_r(\textbf{r})=0$ in order to avoid power losses due to the propagation of the field inside the lens. As far as the choice of the weighting parameters is concerned, we fixed $k_s$ equal to the area of the pixel and $k_f$ equal to the inverse of the area of the lens normalized to the square amplitude of the considered wavelength.\\
In order to keep under control the overall process, we computed the total field corresponding to the synthesized continuous profile and to the synthesized value of $\tau$ ($\tau=[-2.6084 - j7.8221, 1.7136 + j8.1632]$). The resulting far field patterns are reported in Fig. 3(c) and 3(d) (see dot-dashed blue lines).
As it can be seen, the synthesized GRIN lens well satisfies the far field mask constraints and it keeps almost unchanged the beamwidth (BW) of the main lobes and the sidelobes level (SLL) \cite{balanis_antenna}. \\
Then, starting from the GRIN lens, we adopted the MG theory by considering $11$ rings of rods made up with $SiO_2$ material ($\varepsilon_r=4.5$). In particular, the mixing formula has been applied by sampling the permittivity distribution in the center of the rods, which are arranged as in \cite{sun2014luneburg}. Fig. 3(b) shows the obtained $GPC_{F}$, while the corresponding far field patterns are reported in Fig. 3(c) and 3(d) (see dotted red lines).
\begin{figure*}[h!]
\centering
\subfigure[]{\includegraphics[scale=0.15]{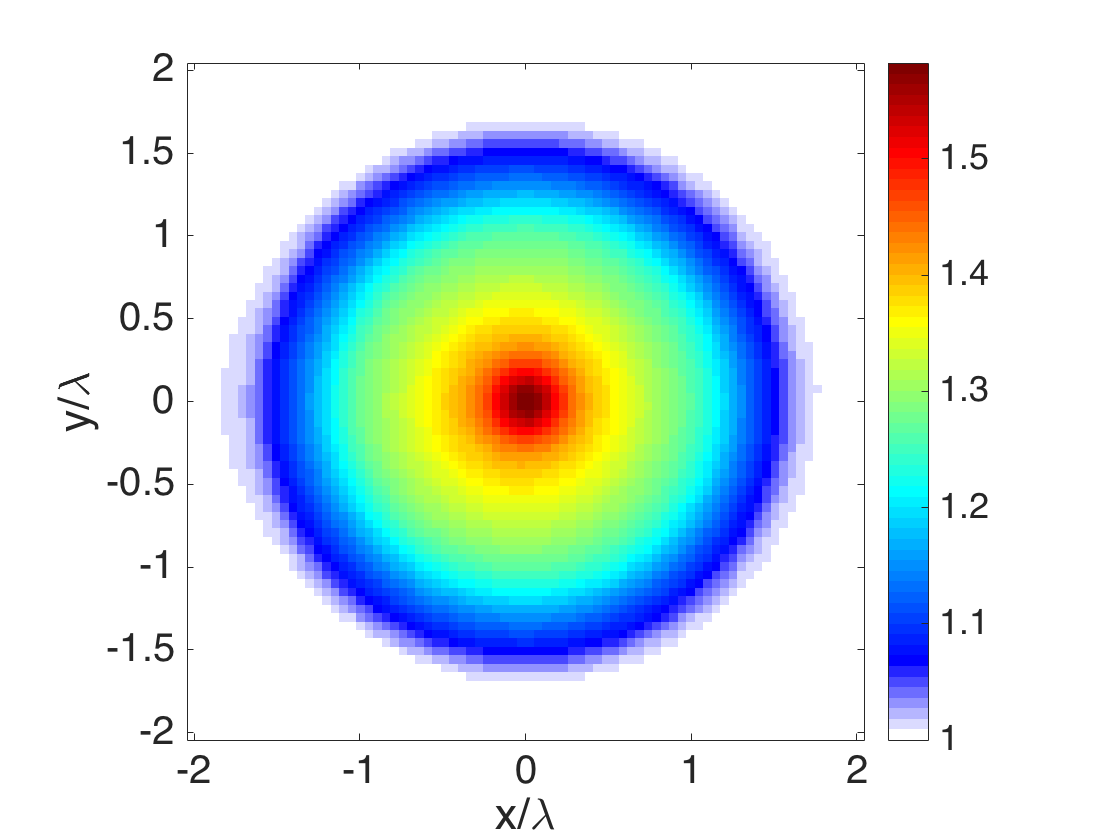} \label{fig:es_cont}}
\subfigure[]{\includegraphics[scale=0.15]{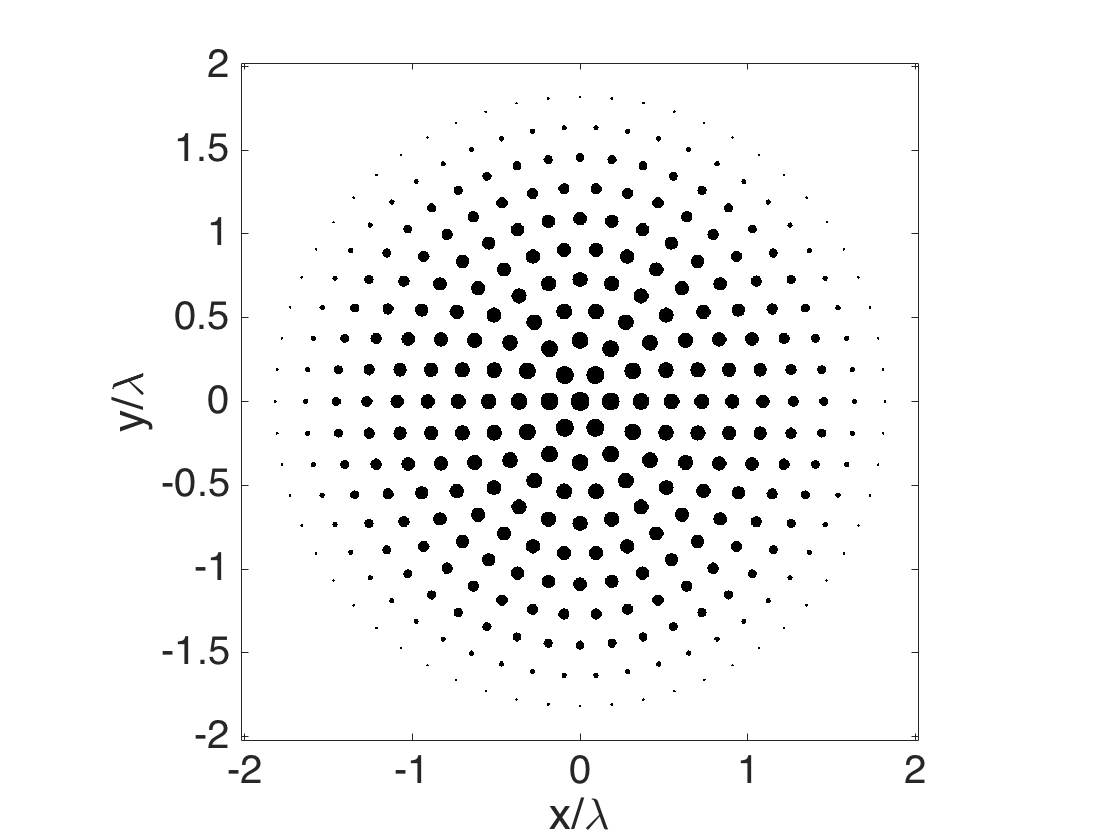}}\\\label{fig:es_MG}
\subfigure[]{\includegraphics[scale=0.30]{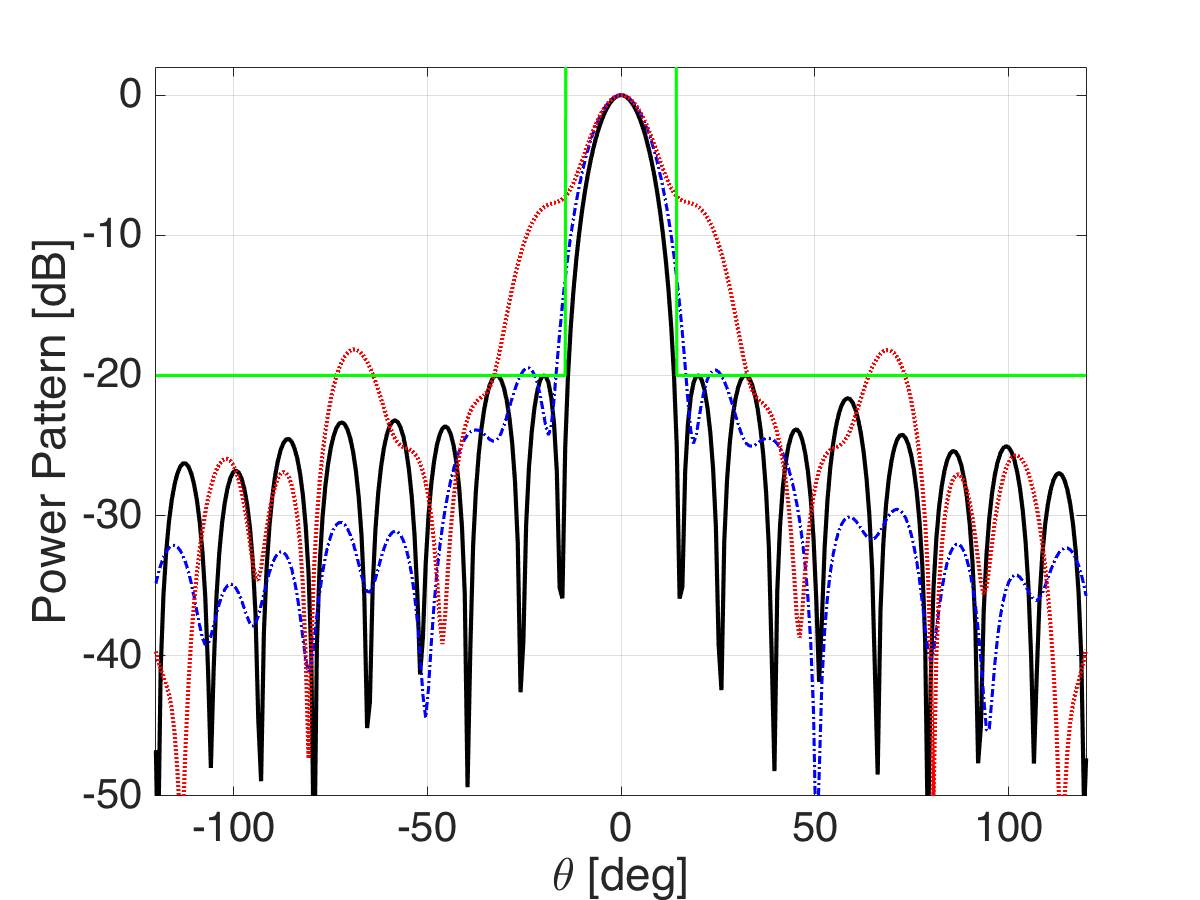}}\label{fig:es_somma1}
\subfigure[]{\includegraphics[scale=0.30]{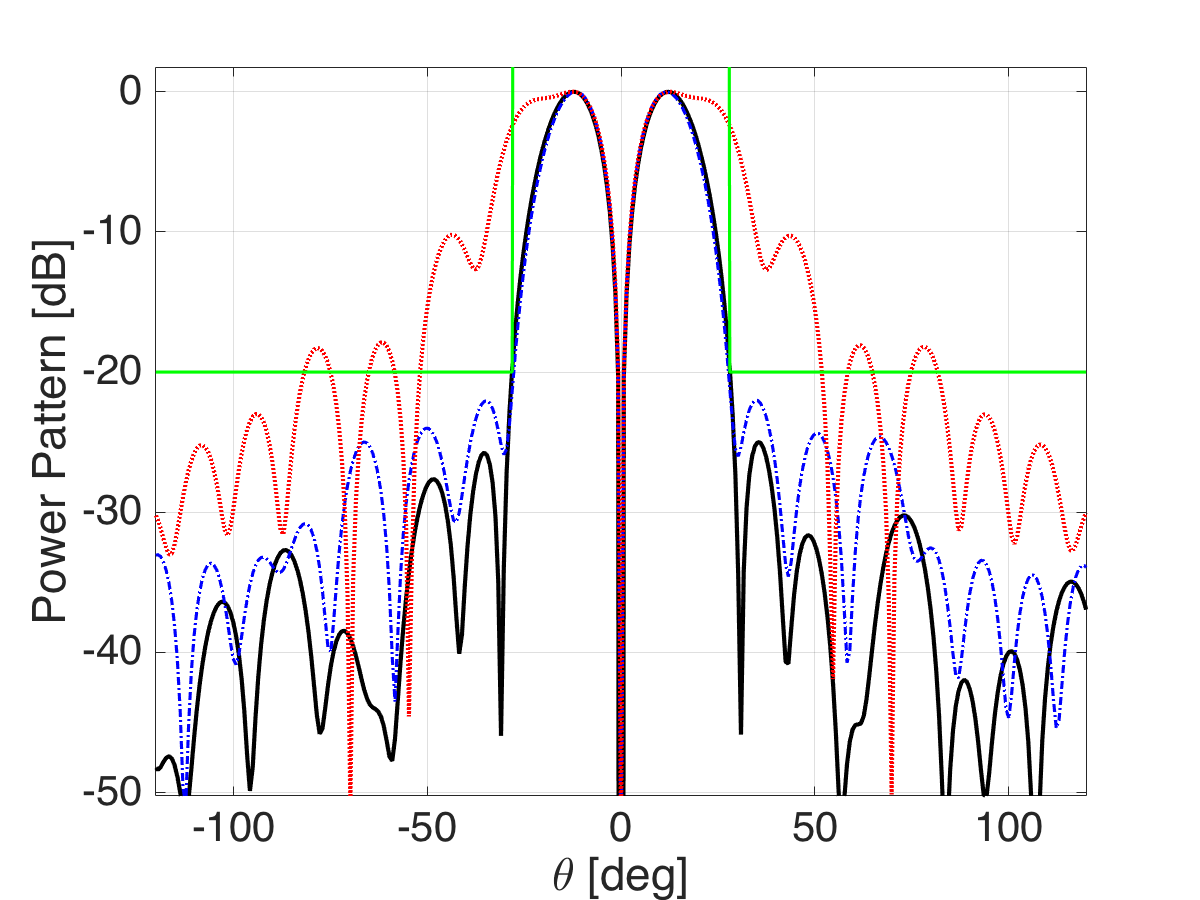}}\label{fig:es_delta1}
\caption{Real part of the permittivity function of the (a) GRIN lens by inverse scattering ($k_s=2.5\cdot 10^{-3}$, $k_f=0.1$, $N=80$) and (b) equivalent GPCs lens by MG ($\varepsilon_r=4.5$, $N=1224$). Far field (c) sum and (d) difference power patterns radiated by profile-(a) (dot-dashed blue lines) and profile-(b) (dotted red lines). Continuous black and green lines represent the specified far field power patterns and the mask constraints, respectively.}
\end{figure*}\label{fig:sintesi}
As it can be seen, the first synthesized $GPC_{F}$-based antenna does not fulfill expectations, since its patterns do not match the given ones and the mask constraints are also violated. This is probably due to the rapidly varying profile of the reflective index, which compromises the performance of the homogenization. Along this line of reasoning, we solved again the ISP by considering in the CSI functional \eqref{func} a further penalty term enforcing a smooth radial variation of $\chi$, i.e., $\left \| \frac{\partial \chi}{\partial r}  \right \|_2^2$. Interestingly, 
the application of the homogenization to the thus obtained GRIN profile leads to a much better solution. The GRIN lens with a smooth profile and the equivalent $GPC_{F}$ are shown in Fig. 4(a) and 4(b), while the corresponding far field patterns are reported in Fig. 4(c) and 4(d) with different colors ($\tau=[-4.5029 - j7.0872,   4.1795 + j6.8097]$). The weighting parameter for the new penalty term has been chosen in such a way that it changes its value iteratively on the basis of intermediate results, as also proposed in \cite{candes2008}. Such a choice, supported by an extensively numerical analysis, has lead to the best radative performances of the equivalent $GPC_{F}$. \\
As a further analysis, we performed the simulation by setting the material of the rods to a lower permittivity value $\varepsilon_r=1.8$ and we observed that the homogenization gives better performances. However, this value of permittivity does not correspond to commonly used materials in the GPCs manufacturing.
All the above statements are supported by the value of the synthetic parameters (SLL and BW) reported in Table \ref{tab:param}.
\begin{figure*}
\centering
\subfigure[]{\includegraphics[scale=0.15]{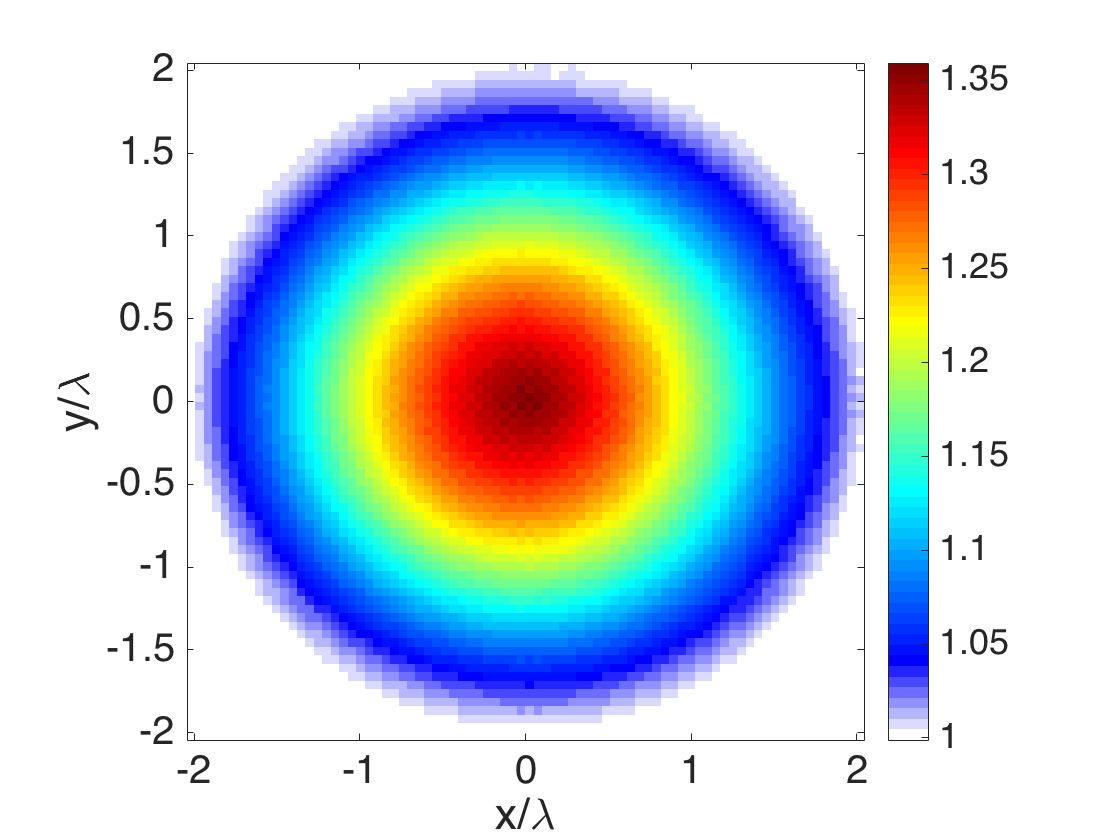}} \label{fig:es_cont}
\subfigure[]{\includegraphics[scale=0.15]{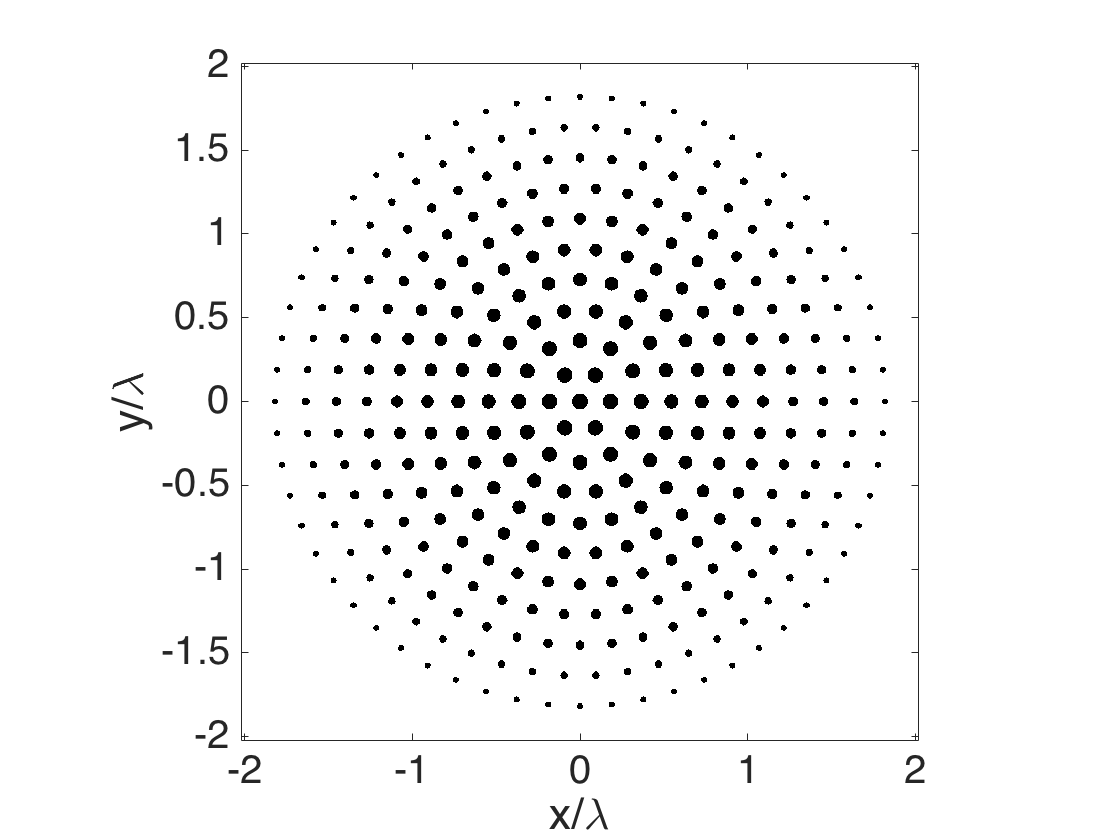}}\\\label{fig:es_MG}
\subfigure[]{\includegraphics[scale=0.30]{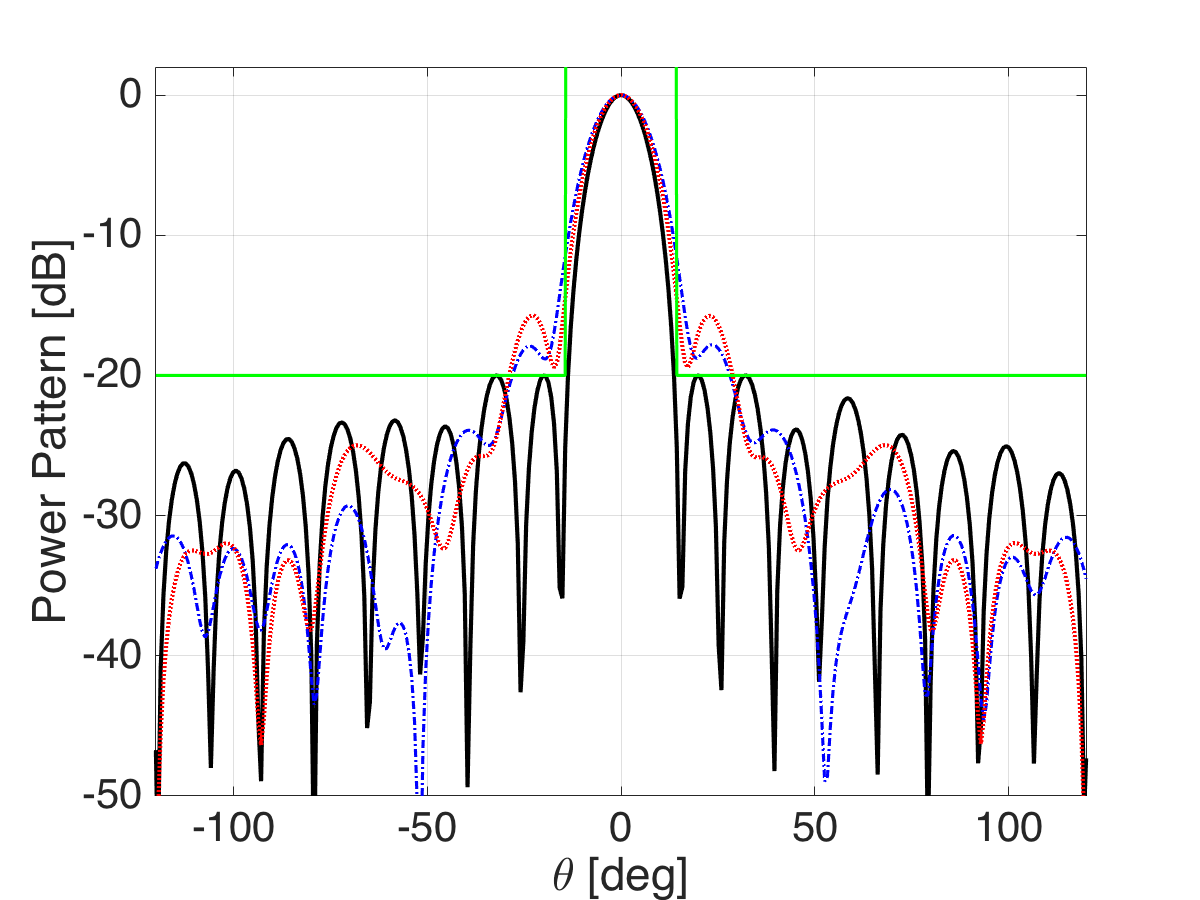}}\label{fig:es_somma1}
\subfigure[]{\includegraphics[scale=0.30]{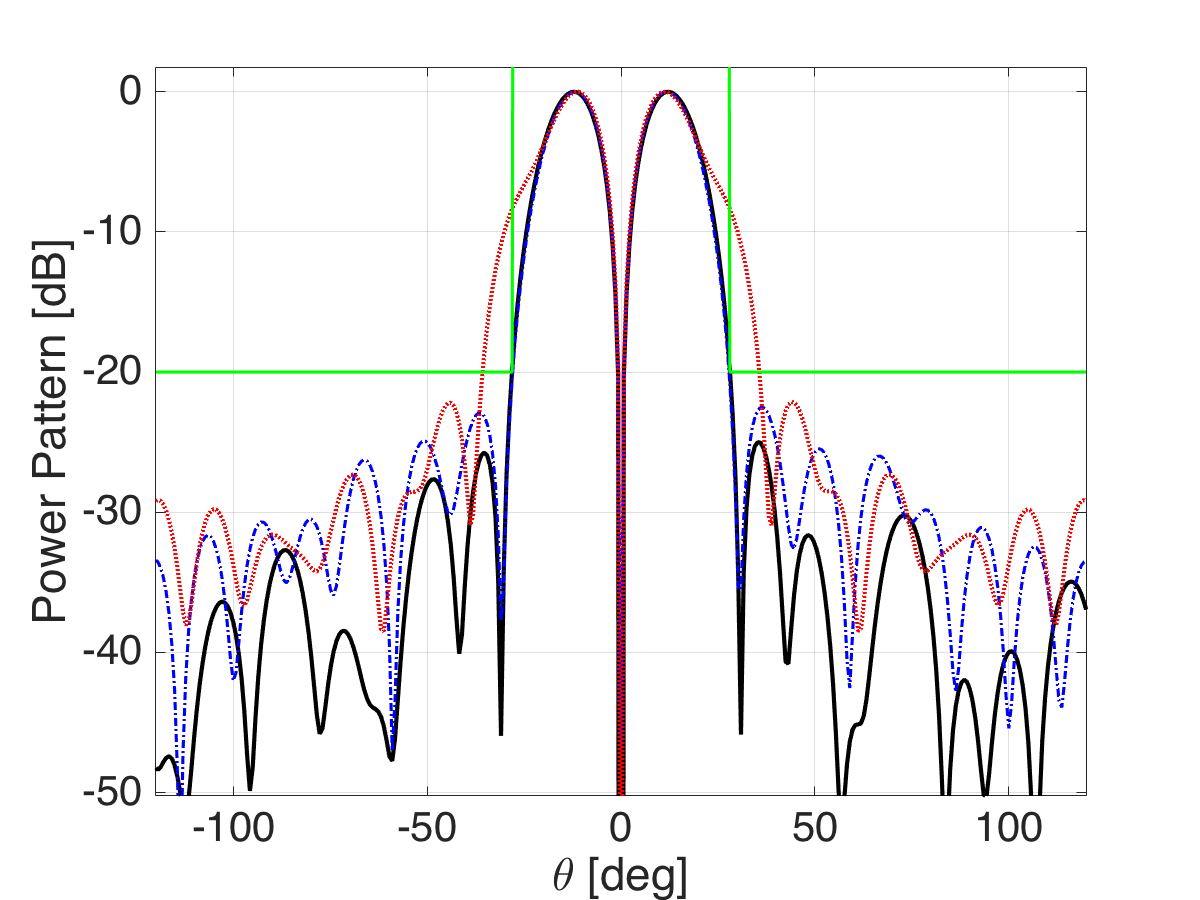}}\label{fig:es_delta1}
\caption{Real part of the permittivity function of the (a) GRIN lens by inverse scattering with smooth constraint ($k_s=2.5\cdot 10^{-3}$, $k_f=0.1$, $N=80$) and (b) equivalent GPCs lens by MG ($\varepsilon_r=4.5$, $N=1224$). Far field (c) sum and (d) difference power patterns radiated by profile-(a) (dot-dashed blue lines) and profile-(b) (dotted red lines). Continuous black and green lines represent the specified far field power patterns and the mask constraints, respectively.}
\end{figure*}\label{fig:sintesi}

In order to test the strategy introduced in Section \ref{sec:rods}, we first solved the modified CSI by considering the expansion \eqref{chi_basis} and using $K=11$ and $a=\lambda/15$, as well as the same arrangement of rods as in \cite{sun2014luneburg}. Also in this case, the penalty term $\Phi_f$ has been added to the functional in order to deal with natural materials. Fig. 5(a) shows the permittivity profile of the obtained $GPC_{R}$-based lens. Note that the synthesis procedure leads to a permittivity value for the external ring equal to 1, so that the overall device results smaller (and actually composed by 10 rings). By using the synthesized value of $\tau$ ($\tau=[0.8348 - j8.5656,  -1.7077 + j8.7196]$), one achieves the corresponding $\Sigma$ and $\Delta$ far field patterns, which are depicted in Figs. 5(c) and 5(d) (dot-dashed blue lines). As it can be seen, the new strategy that allows to directly synthesize $GPC_{R}$ works well, since the design constraints as well as the far fields masks are fulfilled. \\
Finally, we applied the analytic interchange procedure summarized in Fig. 2, by still considering a dielectric material with $\varepsilon_r=4.5$ for the inclusions. The thus obtained $GPC_{F}$ profile is shown in Fig. 5(b), and the corresponding fields in Figs. 5(c) and 5(d) (red dotted lines). As it can be observed, the resulting gradient of the filling factor allows to control the electromagnetic field path and to fully satisfy the initial specifications. 

As a final test aimed to further reduce the complexity of the synthesized devices (as well as to further compare the two introduced strategies), we tried to reduce the number of rings. Hence, with reference to the second approach, we run the minimization of \eqref{func} by using $K=9$ (and $a=\lambda/10$).
In Fig. 6(a) the corresponding $GPC_{R}$ lens is reported. Note that also in this case the rods in the outer ring have unitary permittivity and hence the lens is actually composed by 8 rings. As it can be seen from the corresponding patterns in Fig. 6(c) and 6(d) (dot-dashed blue lines), it still allows to perform at best the pattern reconfiguration ($\tau=[0.2553 - j8.4487,  -1.2238 + j8.7385]$). Interestingly, also the analytical interchanging is successful, so that an effective $GPC_F$ antenna is finally achieved (see Fig. 6(b) and the dotted red lines in Fig. 6(c) and 6(d)). Notably (see also Table \ref{tab:param}), even looking for a smooth profile in the first step, the implementation of the first proposed strategy using homogenization procedure (and exploiting the same number of rings) fails. 

As a consequence of such an example (and many others) it can be concluded that the strategy of Section \ref{sec:rods} outperforms the more straightforward procedure of Section \ref{sec:MG}. This can be attributed to the circumstance of avoiding in the second strategy the intermediate synthesis of a continuous profile, whose characteristics may be difficult to emulate by means of a GPCs. 
\begin{figure*}
\centering
\subfigure[]{\includegraphics[scale=0.15]{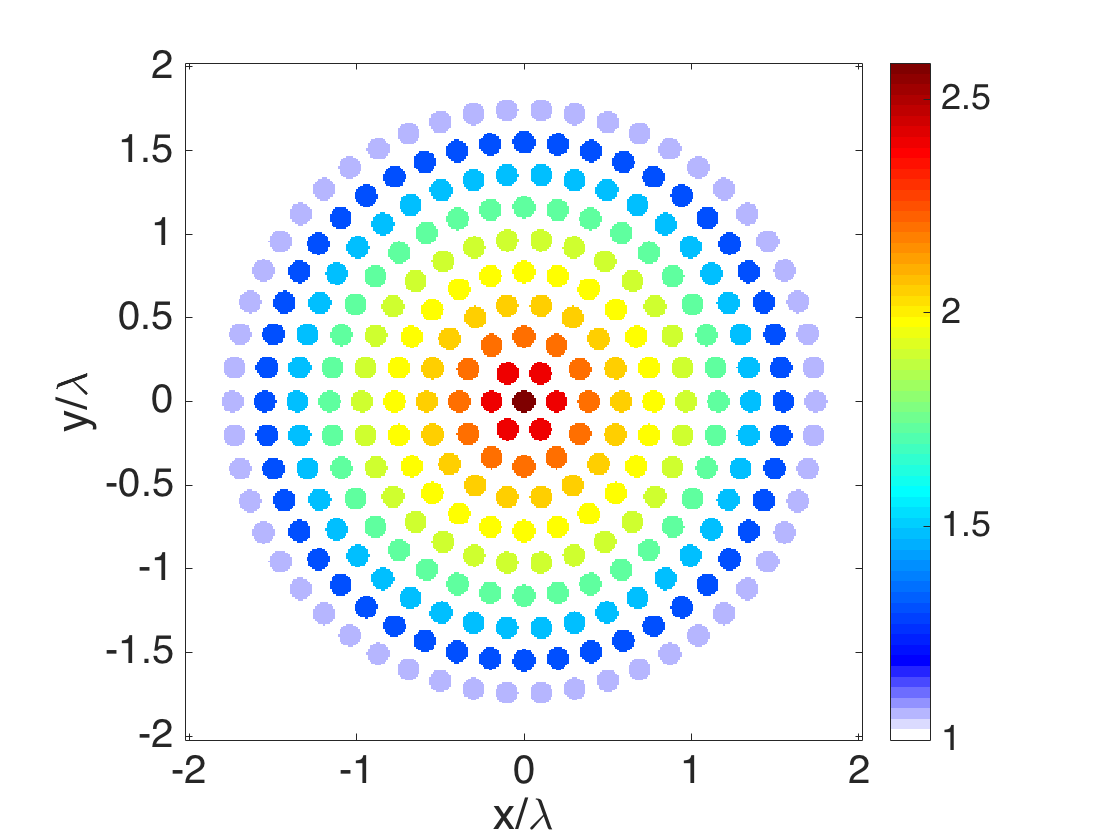}} \label{fig:es_cil}
\subfigure[]{\includegraphics[scale=0.15]{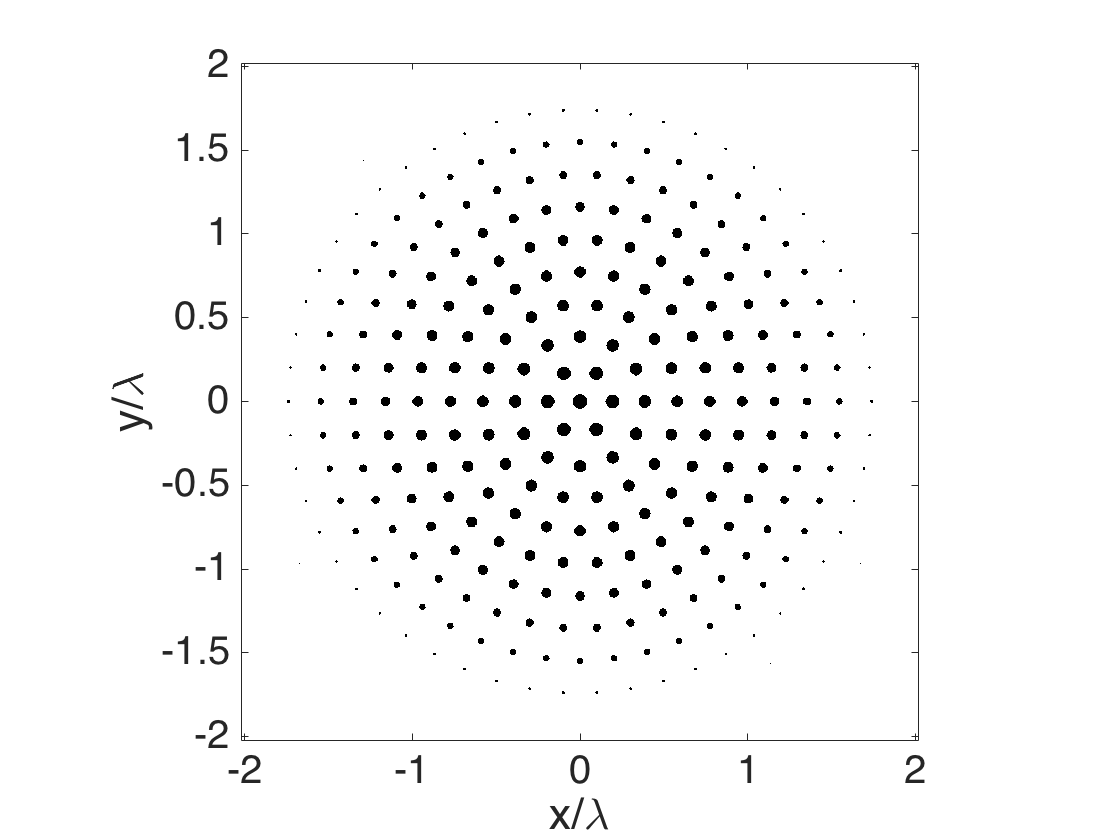}}\\\label{fig:es_rods}
\subfigure[]{\includegraphics[scale=0.30]{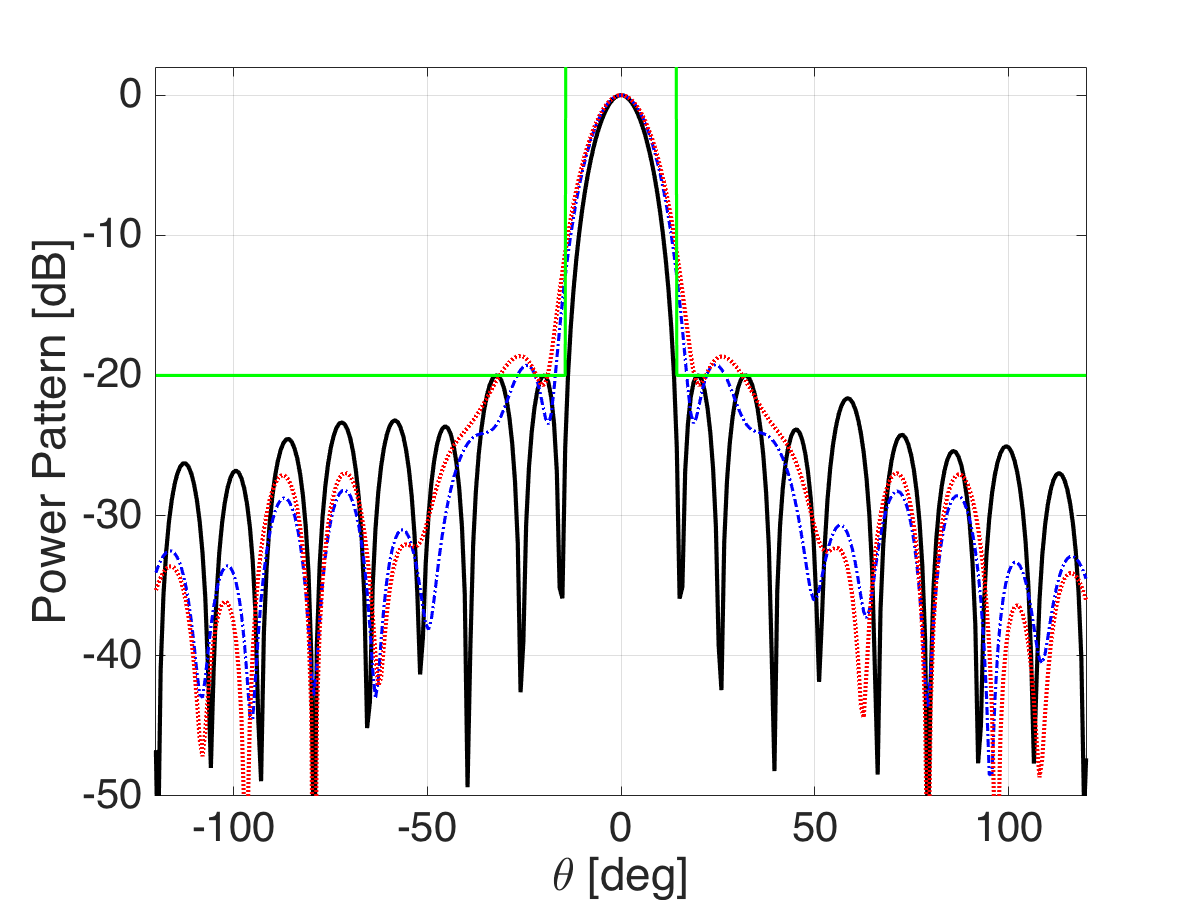}}\label{fig:es_somma2}
\subfigure[]{\includegraphics[scale=0.30]{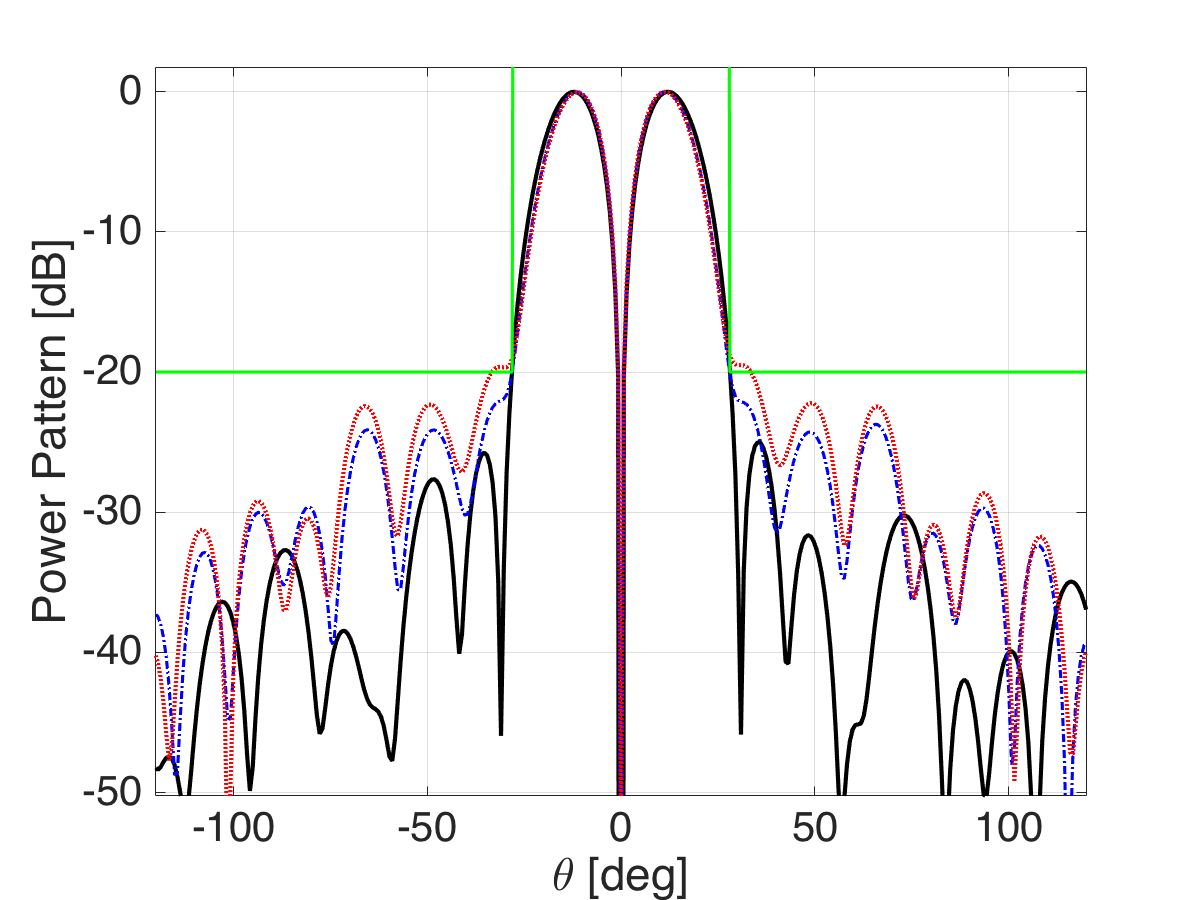}}\label{fig:es_delta2}
\caption{Real part of the permittivity function of the (a) graded refractive index GPCs lens by inverse scattering and expansion \eqref{chi_basis} ($N=408$) and (b) equivalent GPCs lens with gradient filling factor by interchanging procedure as in Section \ref{sec:rods} ($\varepsilon_r=4.5$, $N=1224$). Far field (c) sum and (d) difference power patterns radiated by profile-(a) (dot-dashed blue lines) and profile-(b) (dotted red lines). Continuous black and green lines represent the specified far field power patterns and the mask constraints, respectively. }
\end{figure*}
\begin{figure*}
\centering
\subfigure[]{\includegraphics[scale=0.15]{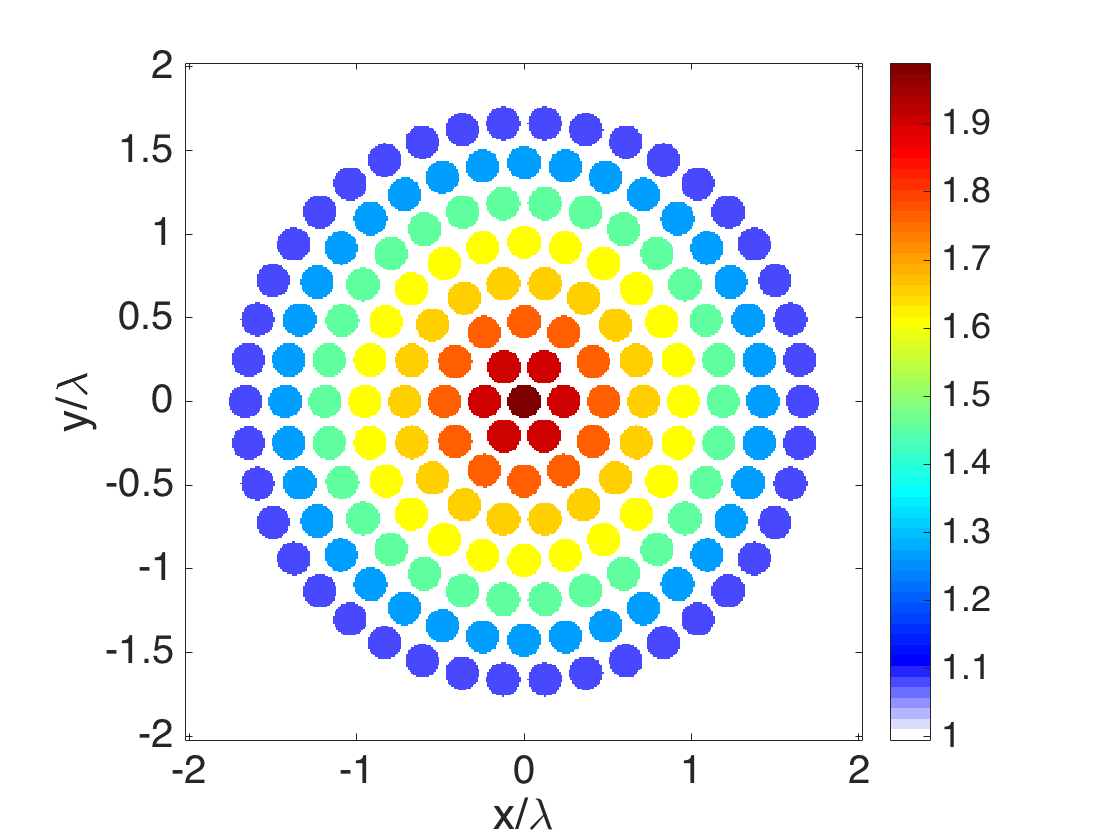}} \label{fig:es_cil}
\subfigure[]{\includegraphics[scale=0.15]{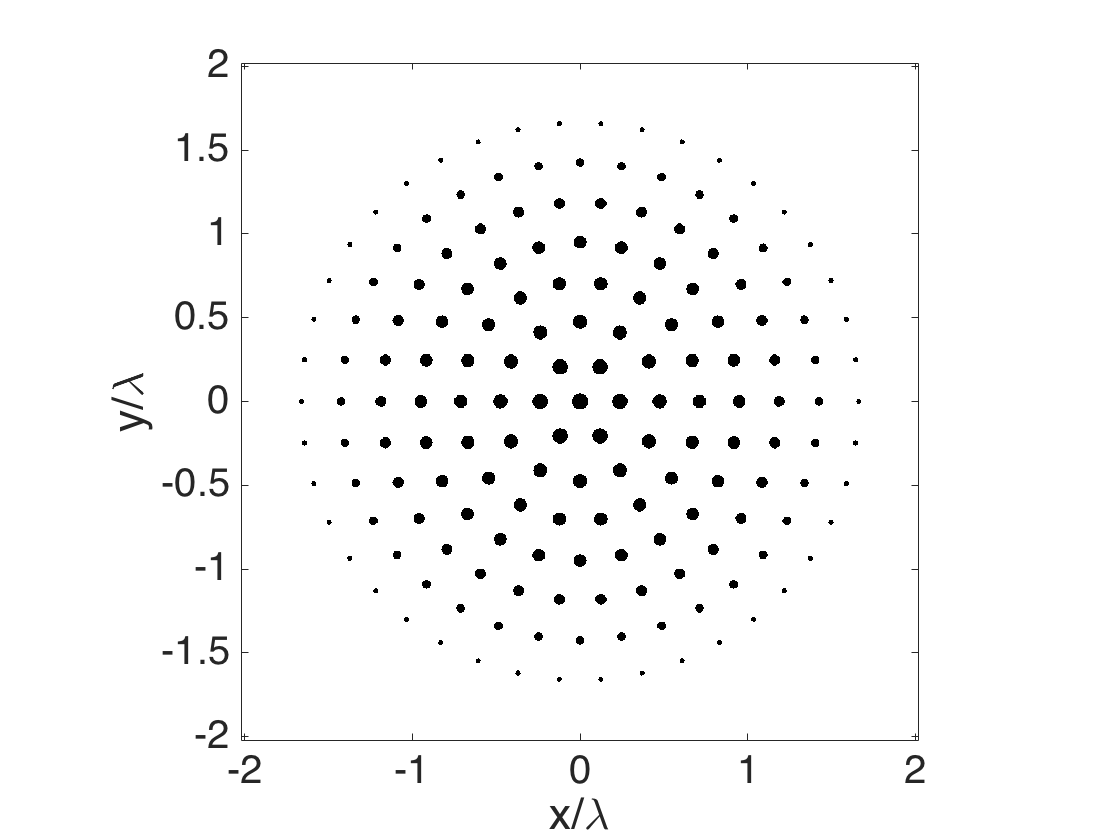}}\\\label{fig:es_rods}
\subfigure[]{\includegraphics[scale=0.30]{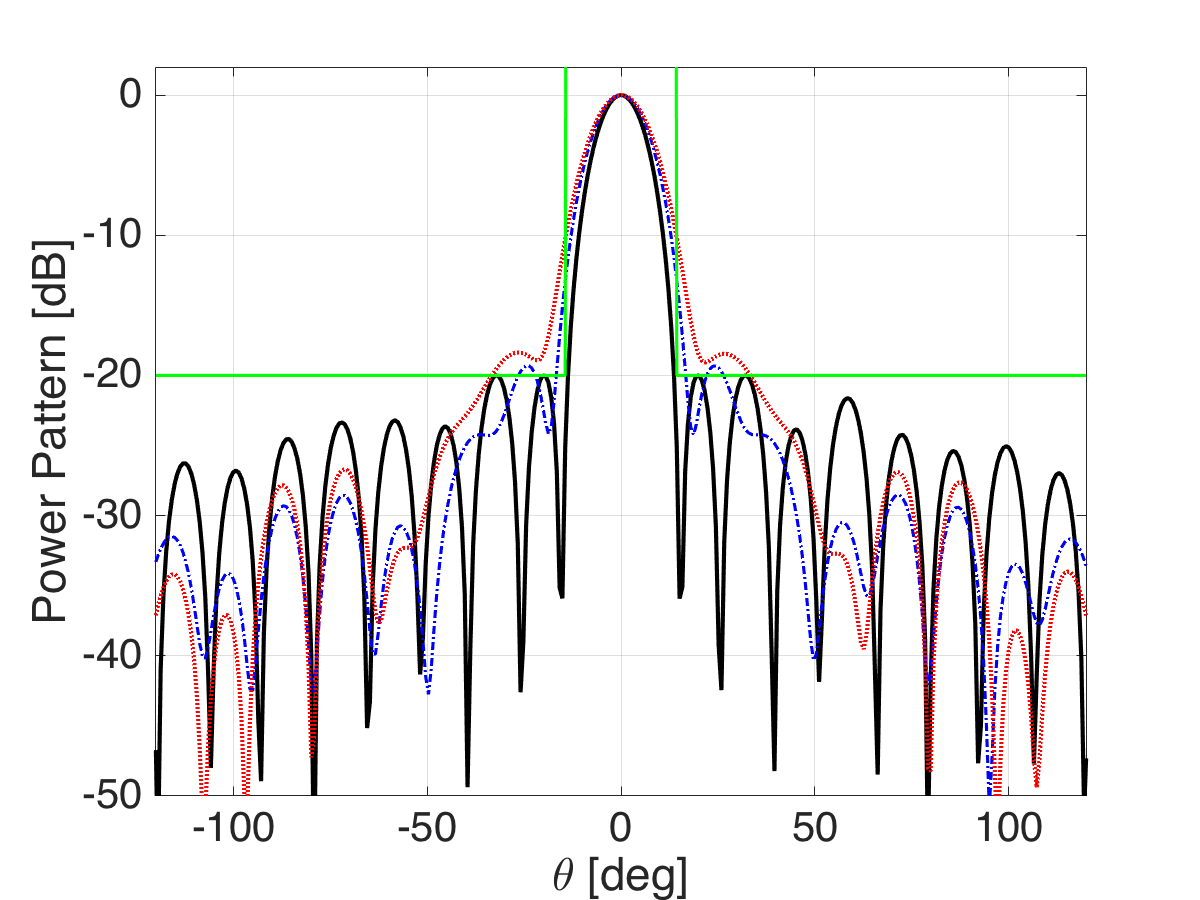}}\label{fig:es_somma2}
\subfigure[]{\includegraphics[scale=0.30]{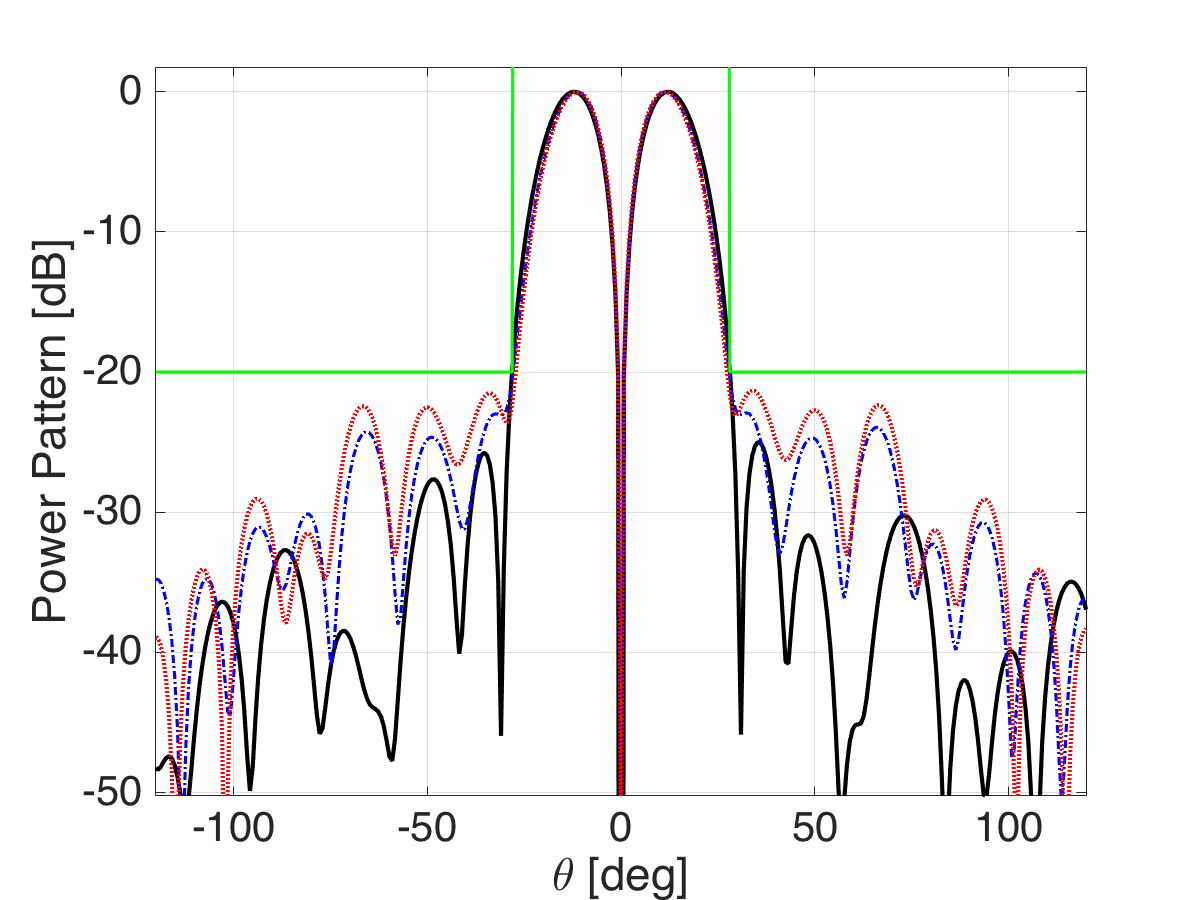}}\label{fig:es_delta2}
\caption{Real part of the permittivity function of the (a) graded refractive index GPCs lens by inverse scattering and expansion \eqref{chi_basis} with a reduced number of rings ($N=408$) and (b) equivalent GPCs lens with gradient filling factor by interchanging procedure. Far field (c) sum and (d) difference power patterns radiated by profile-(a) (dot-dashed blue lines) and profile-(b) (dotted red lines). Continuous black and green lines represent the specified far field power patterns and the mask constraints, respectively.  }
\end{figure*}
\begin{table*}[htbp]\footnotesize
\centering
\caption{\bf Comparison of the synthetic parameters for the far field patterns.}
\label{tab:param}
\renewcommand\arraystretch{1.3}
\begin{tabular}{lcccc}
\multicolumn{1}{c}{\textbf{}} & \multicolumn{2}{c}{\textbf{BW @ -20dB {[}deg{]}}} & \multicolumn{2}{c}{\textbf{SLL {[}dB{]}}} \\ \hline \hline
\multicolumn{1}{c}{}                & \textbf{$\Sigma$}       & \textbf{$\Delta$}       & \textbf{$\Sigma$}         & \textbf{$\Delta$}         \\ \cline{2-5} 
\textbf{mask constraints}                       & 28                   & 56                    & -20                       & -20                    \\ \hline
\textbf{Case 1}                       &                    &                    &                      &                     \\ \hline
\textit{(a)} GRIN by inverse scattering (IS)          & 33                    & 56                    & -19.6                     & -22.03                    \\
\textit{(b)} $GPC_{F}$ from \textit{(a)} by MG ($K=11$, $\varepsilon=4.5$)               & 66                   & 103.6                   & -18.2                     & -18                     \\
\textit{(c)} $GPC_{F}$ from \textit{(a)} by MG ($K=11$, $\varepsilon=1.8$)               & 40                   & 55                   & -18.1                     & -20.3                     \\
\textit{(d)} $GPC_{F}$ from \textit{(a)} by MG ($K=9$, $\varepsilon=4.5$)               & 66                   & 97                   & -18.5                     & -17.8                     \\ \hline
\textit{(e)} GRIN by IS and `smooth' constraint          & 39                    & 56                    & -17.8                     & -22.5                   \\
\textit{(f)} $GPC_{F}$ from \textit{(e)} by MG ($K=11$, $\varepsilon=4.5$)               & 34.5                   & 70                   & -15.75                     & -22.15           \\         
\textit{(g)} $GPC_{F}$ from \textit{(e)} by MG ($K=9$, $\varepsilon=4.5$)               & 56                   & 72                   & -15.1                     & -19.6                      \\ \hline
\textbf{Case 2}                       &                    &                    &                      &                     \\ \hline
\textit{(h)} $GPC_{R}$ by IS ($K=11$)                       & 34                   & 56                    & -19.3                       & -23.7                      \\
\textit{(i)} $GPC_{F}$ from \textit{(h)} by analyt. tool ($\varepsilon=4.5$)                    & 37.5                    & 66                      & -18.65                       & -22.45                     \\ \hline 
\textit{(j)} $GPC_{R}$ by IS ($K=9$)                       & 34                   & 56                    & -19.3                       & -23.9                     \\
\textit{(k)} $GPC_{F}$ from \textit{(j)} by analyt. tool ($\varepsilon=4.5$)                    & 45                    & 55                      & -18.45                       & -21.3                     \\ \hline 
\end{tabular}
\end{table*}

\section{Conclusion}
A new innovative tool has been proposed for the design of two-dimensional graded index Photonic Crystals (GPCs) devices, starting from the solution of an inverse scattering problem. To this end, we have first recast the Contrast Source Inversion method to adjust the amplitude of the primary source, as well as to promote some behavior of the contrast function. 
A first strategy to design GPCs with graded filling factor ($GPC_{F}$) has been introduced by exploiting the widely-used homogenization theories. Then, as a second opportunity, a novel expansion for the contrast function has been proposed, which allows to directly look for a GPCs with a gradient of the refractive index ($GPC_{R}$) profile (rather than for a generic graded index profile).
Although these kinds of GPCs based devices are of interest \textit{by per se}, a simple analytical \virgolette{interchanging} procedure has been proposed which allows to achieve effective $GPC_F$ solutions.\\
The different tools have been assessed through the synthesis of a monopulse antenna radiating $\Sigma/\Delta$ reconfigurable patterns. Both the proposed techniques are able to fulfill the specifications, although the second, and more original one, seems to  exhibit better performances.\\
As a final comment, we would like to stress that the proposed approaches and tools are not restricted to the realization of `canonical' fields, and that they can be applied to generic (physically feasible, see \cite{bucci1997dof}) field specifications, as well as using other unit cells for the definition of the GPCs structure.\\
According to the above encouraging results, possible extension to 3D structures as well as to other kinds of devices is worth to be pursued.

\section*{APPENDIX: Synthesis of the $\Sigma$/$\Delta$ pattern}\label{}

The aim of the field synthesis procedure is the determination of the target fields $E_t$ on the observation points $\textbf{r}_o\in \Gamma_o$ in the near field region, such to fulfill given mask constraints and specifications on the corresponding far fields. 
To this end, let us consider an expansion of the $\Sigma$-field in circular harmonics for a 2D TM polarization:
\begin{equation}\label{eq:ap1}
E_\Sigma \left(\theta,r_{o_{FF}} \right)=\sum_{n=-\lceil k_b R\rceil}^{\lceil k_b R \rceil} \widehat{\gamma}_n e^{jn\theta} ,
\end{equation}
where $\theta$ denotes the angular variable, $R$ is the radius of the lens, $r_{o_{FF}}$ is the radius of a far field observation circle, and:
\begin{equation}\label{eq:ap2}
\widehat{\gamma}_n =\gamma_n H_n^{(2)}\left( k_b r_{o_{FF}} \right) ,
\end{equation}
are the expansion's coefficients. In eq.\eqref{eq:ap1}, the summation has been limited to $-\lceil k_b R\rceil$, $\lceil k_b R\rceil$ in accordance with the finite number of degrees of freedom associated to a source enclosed in a circle of radius $R$ \cite{bucci1997dof}.\\
Then, the $\Delta$-field can be expressed as a linear combination of two $\Sigma$-fields shifted by $\theta_t$, namely:

\begin{equation}\label{eq:ap3}
\begin{split}
E_\Delta \left(\theta,r_{o_{FF}} \right) = E_\Sigma \left(\theta-\theta_t, r_{o_{FF}} \right) - E_\Sigma \left(\theta+\theta_t, r_{o_{FF}} \right) \\
=\sum_{n=-\lceil k_b R\rceil}^{\lceil k_b R \rceil} \widehat{\gamma}_n \left[ e^{jn(\theta - \theta_t)}-e^{jn(\theta + \theta_t)} \right] \\
=\sum_{n=-\lceil k_b R\rceil}^{\lceil k_b R \rceil} \widehat{\gamma}_n \left[ e^{-jn\theta_t}-e^{jn\theta_t} \right] e^{jn\theta}=\sum_{n=-\lceil k_b R\rceil}^{\lceil k_b R \rceil} \widehat{\delta}_n e^{jn\theta}  , 
\end{split}
\end{equation}
where \begin{math} \widehat{\delta}_n=-2j\sin(n\theta_t)\widehat{\gamma}_n \end{math} . 
By so doing, only one between the $\widehat{\gamma}_n$ and $\widehat{\delta}_n$ sets of coefficients has to be evaluated in order to perform the synthesis (the other one being related to it by a simple linear relationship). \\
Then, by noticing that expressions \eqref{eq:ap1} and \eqref{eq:ap3} resemble the expression of uniformly spaced array factors, one can follow the approaches respectively developed in the optimal \virgolette{separate} synthesis of pencil \cite{isernia2000effective} and difference \cite{bucci2005optimal} beams, as well as recent extensions to reconfigurable fields \cite{morabito2010optimal,rocca2015optimal}. By exploiting these results, the unknown coefficients can be finally determined by solving the following Convex Programming problem:
\begin{equation}\label{eq:obiettivo}
\min_{\widehat{\gamma}_n} \biggl\{ -Re  \left[{\left. \frac{dE_\Delta(\theta)}{d\theta} \right |}_{\theta=0}   \right]    \biggr\}  ,
\end{equation}
subject to:
\begin{subnumcases}{}
    Re \left[ {\left. \frac{dE_\Delta(\theta)}{d\theta} \right |}_{\theta=0} \right]   \ge 0  \label{eq:con1}
   \\
   E_\Delta \left(\theta=0\right)=0 \label{eq:con2}
   \\
   \left | E_\Delta(\theta)  \right | ^2 \le UB^\Delta (\theta) \label{eq:con3}
   \\
   Re\left [ \left. E_\Sigma \left(\theta\right) \right |_{\theta=0} \right ] =1\label{eq:con4}
   \\
   Im\left [ \left. E_\Sigma \left(\theta\right) \right |_{\theta=0} \right ] =0\label{eq:con5}
   \\
   \left | E_\Sigma(\theta)  \right | ^2 \le UB^\Sigma (\theta)\label{eq:con6}
\end{subnumcases}
\\
where the objective function \eqref{eq:obiettivo} and constraint \eqref{eq:con1} allow to maximize the amplitude of the (real) derivative of $E_\Delta$ in the target direction, constraints \eqref{eq:con2},\eqref{eq:con4} and \eqref{eq:con5} define the amplitude of the two fields in the target direction, and constraints \eqref{eq:con3} and \eqref{eq:con6} allow to keep under control the sidelobes level of the two power patterns ($UB^\Delta$ and $UB^\Sigma$  being suitable user-defined upper-bound masks).

As far as $\theta_t$ is concerned, it is easy to show that an optimal choice is the first null of the sum pattern. In fact, this allows a physical optimization of the difference pattern slope. 

After solving problem \eqref{eq:obiettivo}-(18), the final expression of the two patterns on $\textbf{r}_o\in\Gamma_o$ located in the near field can be identified by a field backpropagation, i.e.,

\begin{equation}
E_\Sigma \left(\theta,r_o \right)=\sum_{n=-\lceil k_b R\rceil}^{\lceil k_b R \rceil} \gamma_n H_n^{(2)}\left( k_b,r_o \right) e^{jn\theta} ,
\end{equation}

\begin{equation}
E_\Delta \left(\theta,r_o \right) = E_\Sigma \left(\theta-\theta_t,r_o \right) - E_\Sigma \left(\theta+\theta_t,r_o \right) ,
\end{equation}
wherein, by virtue of \eqref{eq:ap2}, it is \begin{math}  \gamma_n=\widehat{\gamma}_n/H_n^{(2)}\left( k_b,r_{o_{FF}} \right) \end{math}.

\footnotesize
\bibliographystyle{IEEEtran}
\bibliography{refs.bib}

\end{document}